# Impact of cross-section uncertainties on supernova neutrino spectral parameter fitting in the Deep Underground Neutrino Experiment

A. Abed Abud,[34] B. Abi,[160] R. Acciarri,[66] M. A. Acero,[11] M. R. Adames,[199] G. Adamov,[72] M. Adamowski,[66] D. Adams,[19] M. Adinolfi,[18] C. Adriano,[29] A. Aduszkiewicz,[80] J. Aguilar,[130] Z. Ahmad,[211] J. Ahmed,[214] B. Aimard,[52] F. Akbar,[179] K. Allison,[42] S. Alonso Monsalve,[34] M. Alrashed,[122] A. Alton,[12] R. Alvarez,[38] P. Amedo,[85,84] J. Anderson,[7] D. A. Andrade,[87] C. Andreopoulos,[182,132] M. Andreotti,[94,67] M. P. Andrews,[66] F. Andrianala,[5] S. Andringa,[131] N. Anfimov,[120] W. L. Anicézio Campanelli,[62] A. Ankowski,[189] M. Antoniassi,[199] M. Antonova,[84] A. Antoshkin,[120] A. Aranda-Fernandez,[41] L. Arellano,[138] L. O. Arnold,[44] M. A. Arroyave,[59] J. Asaadi,[203] A. Ashkenazi,[200] L. Asquith,[197] E. Atkin,[88] D. Auguste,[164] A. Aurisano,[39] V. Aushev,[128] D. Autiero,[111] M. Ayala-Torres,[40] F. Azfar,[160] A. Back,[91] H. Back,[161] J. J. Back,[214] I. Bagaturia,[72] L. Bagby,[66] N. Balashov,[120] S. Balasubramanian,[66] P. Baldi,[23] W. Baldini,[94] B. Baller,[66] B. Bambah,[81] R. Banerjee,[221] F. Barao,[131,113] G. Barenboim,[84] P. Barham Alzás,[34] G. J. Barker,[214] W. Barkhouse,[152] C. Barnes,[142] G. Barr,[160] J. Barranco Monarca,[77] A. Barros,[199] N. Barros,[131,61] J. L. Barrow,[139] A. Basharina-Freshville,[209] A. Bashyal,[7] V. Basque,[66] C. Batchelor,[58] J. B. R. Battat,[215] F. Battisti,[160] F. Bay,[4] M. C. Q. Bazetto,[29] J. L. L. Bazo Alba,[173] J. F. Beacom,[158] E. Bechetoille,[111] B. Behera,[68] E. Belchior,[134] G. Bell,[53] L. Bellantoni,[66] G. Bellettini,[103,171] V. Bellini,[93,30] O. Beltramello,[34] N. Benekos,[34] C. Benitez Montiel,[84,9] D. Benjamin,[19] F. Bento Neves,[131] J. Berger,[43] S. Berkman,[66] P. Bernardini,[97,183] R. M. Berner,[13] A. Bersani,[96] S. Bertolucci,[92,16] M. Betancourt,[66] A. Betancur Rodríguez,[59] A. Bevan,[176] Y. Bezawada,[22] A. T. Bezerra,[62] T. J. Bezerra,[197] J. Bhambure,[194] A. Bhardwaj,[134] V. Bhatnagar,[163] M. Bhattacharjee,[89] M. Bhattacharya,[66] D. Bhattarai,[148] S. Bhuller,[18] B. Bhuyan,[89] S. Biagi,[105] J. Bian,[23] K. Biery,[66] B. Bilki,[14,109] M. Bishai,[19] A. Bitadze,[138] A. Blake,[129] F. D. Blaszczyk,[66] G. C. Blazey,[153] D. Blend,[109] E. Blucher,[36] J. Boissevain,[133] S. Bolognesi,[33] T. Bolton,[122] L. Bomben,[98,108] M. Bonesini,[98,144] C. Bonilla-Diaz,[31] F. Bonini,[19] A. Booth,[176] F. Boran,[14] S. Bordoni,[34] A. Borkum,[197] N. Bostan,[109] P. Bour,[49] J. Bracinik,[15] D. Braga,[66] D. Brailsford,[129] A. Branca,[98] A. Brandt,[203] M. Bravo-Moreno,[73] J. Bremer,[34] C. Brew,[182] S. J. Brice,[66] V. Brio,[93] C. Brizzolari,[98,144] C. Bromberg,[143] J. Brooke,[18] A. Bross,[66] G. Brunetti,[98,144] M. Brunetti,[214] N. Buchanan,[43] H. Budd,[179] J. Buergi,[13] G. Caceres V.,[22] I. Cagnoli,[92,16] T. Cai,[221] D. Caiulo,[111] R. Calabrese,[94,67] P. Calafiura,[130] J. Calcutt,[159] M. Calin,[20] L. Calivers,[13] S. Calvez,[43] E. Calvo,[38] A. Caminata,[96] A. Campos Benitez,[212] D. Caratelli,[26] D. Carber,[43] J. M. Carceller,[209] G. Carini,[19] B. Carlus,[111] M. F. Carneiro,[19] P. Carniti,[98] I. Caro Terrazas,[43] H. Carranza,[203] N. Carrara,[22] L. Carroll,[122] T. Carroll,[218] A. Carter,[180] J. F. Castaño Forero,[6] A. Castillo,[187] C. Castromonte,[106] E. Catano-Mur,[217] C. Cattadori,[98] F. Cavalier,[164] G. Cavallaro,[98] F. Cavanna,[66] S. Centro,[162] G. Cerati,[66] A. Cervelli,[92] A. Cervera Villanueva,[84] K. Chakraborty,[170] M. Chalifour,[34] A. Chappell,[214] E. Chardonnet,[165] N. Charitonidis,[34] A. Chatterjee,[172] S. Chattopadhyay,[211] H. Chen,[19] M. Chen,[23] Y. Chen,[13,189] Z. Chen-Wishart,[180] Y. Cheon,[208] D. Cherdack,[80] C. Chi,[44] S. Childress,[66] R. Chirco,[87] A. Chiriacescu,[20] N. Chitirasreemadam,[103,171] K. Cho,[125] S. Choate,[153] D. Chokheli,[72] P. S. Chong,[168] B. Chowdhury,[7] A. Christensen,[43] D. Christian,[66] G. Christodoulou,[34] A. Chukanov,[120] M. Chung,[208] E. Church,[161] V. Cicero,[92,16] D. Clapa,[213] P. Clarke,[58] G. Cline,[130] T. E. Coan,[193] A. G. Cocco,[100] J. A. B. Coelho,[165] A. Cohen,[165] J. Collot,[76] E. Conley,[56] J. M. Conrad,[139] M. Convery,[189] P. Cooke,[132] S. Copello,[96] P. Cova,[99,166] C. Cox,[180] L. Cremaldi,[148] L. Cremonesi,[176] J. I. Crespo-Anadón,[38] M. Crisler,[66] E. Cristaldo,[99,9] J. Crnkovic,[66] G. Crone,[209] R. Cross,[129] A. Cudd,[42] C. Cuesta,[38] Y. Cui,[25] D. Cussans,[18] J. Dai,[76] O. Dalager,[23] R. Dallavalle,[165] H. da Motta,[32] Z. A. Dar,[217] R. Darby,[197] L. Da Silva Peres,[65] C. David,[221,66] Q. David,[111] G. S. Davies,[148] S. Davini,[96] J. Dawson,[165] K. De,[203] S. De,[2] R. De Aguiar,[29] P. De Almeida,[29] P. Debbins,[109] I. De Bonis,[52] M. P. Decowski,[150,3] A. de Gouvêa,[154] P. C. De Holanda,[29] I. L. De Icaza Astiz,[197] A. Deisting,[137] P. De Jong,[150,3] A. De la Torre,[38] A. Delbart,[33] V. De Leo,[186,104] D. Delepine,[77] M. Delgado,[98,144] A. Dell'Acqua,[34] N. Delmonte,[99,166] P. De Lurgio,[7] J. R. T. de Mello Neto,[65] D. M. DeMuth,[210] S. Dennis,[28] C. Densham,[182] P. Denton,[19] G. W. Deptuch,[19] A. De Roeck,[34] V. De Romeri,[84] G. De Souza,[29] J. P. Detje,[28] R. Devi,[117] J. Devine,[34] R. Dharmapalan,[79] M. Dias,[207] J. S. Díaz,[91] F. Díaz,[173] F. Di Capua,[100,149] A. Di Domenico,[186,104] S. Di Domizio,[96,71] S. Di Falco,[103] L. Di Giulio,[34] P. Ding,[66] L. Di Noto,[96,71] E. Diociaiuti,[95] C. Distefano,[105] R. Diurba,[13] M. Diwan,[19] Z. Djurcic,[7] D. Doering,[189] S. Dolan,[34] F. Dolek,[14] M. J. Dolinski,[55] D. Domenici,[95] L. Domine,[189] S. Donati,[103,171] Y. Donon,[34] S. Doran,[110] D. Douglas,[143] A. Dragone,[189] F. Drielsma,[189] L. Duarte,[207] D. Duchesneau,[52] K. Duffy,[160,66] K. Dugas,[23] P. Dunne,[88] B. Dutta,[201] H. Duyang,[190] O. Dvornikov,[79] D. A. Dwyer,[130] A. S. Dyshkant,[153] M. Eads,[153] A. Earle,[197] S. Edayath,[110] D. Edmunds,[143] J. Eisch,[66] L. Emberger,[138,140] P. Englezos,[181] A. Ereditato,[219] T. Erjavec,[22] C. O. Escobar,[66] J. J. Evans,[138] E. Ewart,[91] A. C. Ezeribe,[188] K. Fahey,[66] L. Fajt,[34] A. Falcone,[98,144] M. Fani',[133] C. Farnese,[101] Y. Farzan,[112] D. Fedoseev,[120] J. Felix,[77] Y. Feng,[110] E. Fernandez-Martinez,[136] F. Ferraro,[96,71] G. Ferry,[164] L. Fields,[155] P. Filip,[48] A. Filkins,[198] F. Filthaut,[150,177] R. Fine,[133] G. Fiorillo,[100,149] M. Fiorini,[94,67] V. Fischer,[110] R. S. Fitzpatrick,[142] W. Flanagan,[51] B. Fleming,[36,219] S. Fogarty,[43] W. Foreman,[87] J. Fowler,[56] J. Franc,[49] K. Francis,[153] D. Franco,[219] J. Freeman,[66] J. Fried,[19] A. Friedland,[189] S. Fuess,[66] I. K. Furic,[68] K. Furman,[176] A. P. Furmanski,[147] A. Gabrielli,[92,16] A. Gago,[173] H. Gallagher,[206]

A. Gallas,[164] N. Gallice,[99,145] V. Galymov,[111] E. Gamberini,[34] T. Gamble,[188] F. Ganacim,[199] R. Gandhi,[78] S. Ganguly,[66] F. Gao,[172] S. Gao,[19] D. Garcia-Gamez,[73] M. Á. García-Peris,[84] S. Gardiner,[66] D. Gastler,[17] A. Gauch,[13] J. Gauvreau,[157] P. Gauzzi,[186,104] G. Ge,[44] N. Geffroy,[52] B. Gelli,[29] S. Gent,[192] L. Gerlach,[19] Z. Ghorbani-Moghaddam,[96] P. Giammaria,[29] T. Giammaria,[94,67] N. Giangiacomi,[205] D. Gibin,[162,101] I. Gil-Botella,[38] S. Gilligan,[159] A. Gioiosa,[103] S. Giovannella,[95] C. Girerd,[111] A. K. Giri,[90] C. Giugliano,[94] D. Gnani,[130] O. Gogota,[128] S. Gollapinni,[133] K. Gollwitzer,[66] R. A. Gomes,[63] L. V. Gomez Bermeo,[187] L. S. Gomez Fajardo,[187] F. Gonnella,[15] D. Gonzalez-Diaz,[85] M. Gonzalez-Lopez,[136] M. C. Goodman,[7] O. Goodwin,[138] S. Goswami,[170] C. Gotti,[98] J. Goudeau,[134] E. Goudzovski,[15] C. Grace,[130] R. Gran,[146] E. Granados,[77] P. Granger,[165] C. Grant,[17] D. Gratieri,[70] P. Green,[160] S. Greenberg,[21,130] L. Greenler,[218] J. Greer,[18] J. Grenard,[34] W. C. Griffith,[197] F. T. Groetschla,[34] M. Groh,[43] K. Grzelak,[213] W. Gu,[19] V. Guarino,[7] M. Guarise,[94,67] R. Guenette,[138] E. Guerard,[164] M. Guerzoni,[92] D. Guffanti,[98] A. Guglielmi,[101] B. Guo,[190] Y. Guo,[194] A. Gupta,[189] V. Gupta,[150,3] K. K. Guthikonda,[126] D. Gutierrez,[174] P. Guzowski,[138] M. M. Guzzo,[29] S. Gwon,[37] C. Ha,[37] K. Haaf,[66] A. Habig,[146] H. Hadavand,[203] R. Haenni,[13] L. Hagaman,[219] A. Hahn,[66] J. Haiston,[191] P. Hamacher-Baumann,[160] T. Hamernik,[66] P. Hamilton,[88] J. Han,[172] J. Hancock,[15] F. Happacher,[95] D. A. Harris,[221,66] J. Hartnell,[197] T. Hartnett,[182] J. Harton,[43] T. Hasegawa,[124] C. Hasnip,[160] R. Hatcher,[66] K. W. Hatfield,[23] A. Hatzikoutelis,[184] C. Hayes,[91] K. Hayrapetyan,[176] J. Hays,[176] E. Hazen,[17] M. He,[80] A. Heavey,[66] K. M. Heeger,[219] J. Heise,[196] S. Henry,[179] M. A. Hernandez Morquecho,[87] K. Herner,[66] V. Hewes,[39] A. Higuera,[178] C. Hilgenberg,[147] T. Hill,[82] S. J. Hillier,[15] A. Himmel,[66] E. Hinkle,[36] L. R. Hirsch,[199] J. Ho,[54] J. Hoff,[66] A. Holin,[182] T. Holvey,[160] E. Hoppe,[161] G. A. Horton-Smith,[122] M. Hostert,[147] T. Houdy,[164] B. Howard,[66] R. Howell,[179] J. Hoyos Barrios,[141] I. Hristova,[182] M. S. Hronek,[66] J. Huang,[22] R. G. Huang,[130] Z. Hulcher,[189] G. Iles,[88] N. Ilic,[205] A. M. Iliescu,[92] R. Illingworth,[66] G. Ingratta,[92,16] A. Ioannisian,[220] B. Irwin,[147] L. Isenhower,[1] M. Ismerio Oliveira,[65] R. Itay,[189] C. M. Jackson,[161] V. Jain,[2] E. James,[66] W. Jang,[203] B. Jargowsky,[23] F. Jediny,[49] D. Jena,[66] Y. S. Jeong,[37] C. Jesús-Valls,[83] X. Ji,[19] J. Jiang,[194] L. Jiang,[212] A. Jipa,[20] J. H. Jo,[19] F. R. Joaquim,[131,113] W. Johnson,[191] B. Jones,[203] R. Jones,[188] N. Jovancevic,[156] M. Judah,[172] C. K. Jung,[194] T. Junk,[66] Y. Jwa,[44] M. Kabirnezhad,[88] A. Kaboth,[180,182] I. Kadenko,[128] I. Kakorin,[120] A. Kalitkina,[120] D. Kalra,[44] O. Kamer Koseyan,[109] F. Kamiya,[64] D. M. Kaplan,[87] G. Karagiorgi,[44] G. Karaman,[109] A. Karcher,[130] Y. Karyotakis,[52] S. Kasai,[127] S. P. Kasetti,[134] L. Kashur,[43] I. Katsioulas,[15] A. Kauther,[153] N. Kazaryan,[220] E. Kearns,[17] P. T. Keener,[168] K. J. Kelly,[34] E. Kemp,[29] O. Kemularia,[72] Y. Kermaidic,[164] W. Ketchum,[66] S. H. Kettell,[19] M. Khabibullin,[107] N. Khan,[88] A. Khotjantsev,[107] A. Khvedelidze,[72] D. Kim,[201] J. Kim,[179] B. King,[66] B. Kirby,[44] M. Kirby,[66] J. Klein,[168] J. Kleykamp,[148] A. Klustova,[88] T. Kobilarcik,[66] L. Koch,[137] K. Koehler,[218] L. W. Koerner,[80] D. H. Koh,[189] S. Kohn,[21,130] P. P. Koller,[13] L. Kolupaeva,[120] D. Korablev,[120] M. Kordosky,[217] T. Kosc,[76] U. Kose,[34] V. A. Kostelecký,[91] K. Kothekar,[18] I. Kotler,[55] V. Kozhukalov,[120] R. Kralik,[197] L. Kreczko,[18] F. Krennrich,[110] I. Kreslo,[13] W. Kropp,[23] T. Kroupova,[168] S. Kubota,[138] M. Kubu,[34] Y. Kudenko,[107] V. A. Kudryavtsev,[188] S. Kuhlmann,[7] S. Kulagin,[107] J. Kumar,[79] P. Kumar,[188] P. Kunze,[52] R. Kuravi,[130] N. Kurita,[189] C. Kuruppu,[190] V. Kus,[49] T. Kutter,[134] J. Kvasnicka,[48] D. Kwak,[208] T. Labree,[153] A. Lambert,[130] B. J. Land,[168] C. E. Lane,[55] K. Lang,[204] T. Langford,[219] M. Langstaff,[138] F. Lanni,[34] O. Lantwin,[52] J. Larkin,[19] P. Lasorak,[88] D. Last,[168] A. Laundrie,[218] G. Laurenti,[92] A. Lawrence,[130] P. Laycock,[19] I. Lazanu,[20] M. Lazzaroni,[99,145] T. Le,[206] S. Leardini,[85] J. Learned,[79] P. LeBrun,[111] T. LeCompte,[189] C. Lee,[66] V. Legin,[128] G. Lehmann Miotto,[34] R. Lehnert,[91] M. A. Leigui de Oliveira,[64] M. Leitner,[130] L. M. Lepin,[138] S. W. Li,[189] Y. Li,[19] H. Liao,[122] C. S. Lin,[130] S. Lin,[134] D. Lindebaum,[18] R. A. Lineros,[31] J. Ling,[195] A. Lister,[218] B. R. Littlejohn,[87] J. Liu,[23] Y. Liu,[36] S. Lockwitz,[66] T. Loew,[130] M. Lokajicek,[48] I. Lomidze,[72] K. Long,[88] N. López March,[84] T. Lord,[214] J. M. LoSecco,[155] W. C. Louis,[133] X.-G. Lu,[214] K. B. Luk,[21,130] B. Lunday,[168] X. Luo,[26] E. Luppi,[94,67] T. Lux,[83] J. Maalmi,[164] D. MacFarlane,[189] A. A. Machado,[29] P. Machado,[66] C. T. Macias,[91] J. R. Macier,[66] M. MacMahon,[209] A. Maddalena,[75] A. Madera,[34] P. Madigan,[21,130] S. Magill,[7] C. Magueur,[164] K. Mahn,[143] A. Maio,[131,61] A. Major,[56] K. Majumdar,[132] J. A. Maloney,[50] M. Man,[205] G. Mandrioli,[92] R. C. Mandujano,[23] J. Maneira,[131,61] L. Manenti,[209] S. Manly,[179] A. Mann,[206] K. Manolopoulos,[182] M. Manrique Plata,[91] S. Manthey Corchado,[38] V. N. Manyam,[19] M. Marchan,[66] A. Marchionni,[66] W. Marciano,[19] D. Marfatia,[79] C. Mariani,[212] J. Maricic,[79] F. Marinho,[114] A. D. Marino,[42] T. Markiewicz,[189] D. Marsden,[138] M. Marshak,[147] C. M. Marshall,[179] J. Marshall,[214] J. Marteau,[111] J. Martín-Albo,[84] N. Martinez,[122] D. A. Martinez Caicedo,[191] F. Martínez López,[176] P. Martínez Miravé,[84] S. Martynenko,[19] V. Mascagna,[98,108] K. Mason,[206] C. Massari,[98] A. Mastbaum,[181] F. Matichard,[130] S. Matsuno,[79] J. Matthews,[134] C. Mauger,[168] N. Mauri,[92,16] K. Mavrokoridis,[132] I. Mawby,[214] R. Mazza,[98] A. Mazzacane,[66] T. McAskill,[215] E. McCluskey,[66] N. McConkey,[209] K. S. McFarland,[179] C. McGrew,[194] A. McNab,[138] A. Mefodiev,[107] P. Mehta,[118] P. Melas,[10] O. Mena,[84] H. Mendez,[174] P. Mendez,[34] D. P. Méndez,[19] A. Menegolli,[102,167] G. Meng,[101] M. D. Messier,[91] W. Metcalf,[134] M. Mewes,[91] H. Meyer,[216] T. Miao,[66] G. Michna,[192] V. Mikola,[209] R. Milincic,[79] G. Miller,[138] W. Miller,[147] J. Mills,[206] O. Mineev,[107] A. Minotti,[98,144] O. G. Miranda,[40] S. Miryala,[19] S. Miscetti,[95] C. S. Mishra,[66] S. R. Mishra,[190] A. Mislivec,[147] M. Mitchell,[134] D. Mladenov,[34] I. Mocioiu,[169] K. Moffat,[57] A. Mogan,[43] N. Moggi,[92,16] R. Mohanta,[81] T. A. Mohayai,[66] N. Mokhov,[66] J. Molina,[9] L. Molina Bueno,[84] E. Montagna,[92,16] A. Montanari,[92] C. Montanari,[102,66,167] D. Montanari,[66] D. Montanino,[97,183] L. M. Montaño Zetina,[40] S. H. Moon,[208] M. Mooney,[43] A. F. Moor,[28] D. Moreno,[6] L. Morescalchi,[103] D. Moretti,[98] C. Morris,[80] C. Mossey,[66] M. Mote,[134] E. Motuk,[209] C. A. Moura,[64] J. Mousseau,[142]

G. Mouster,[129] W. Mu,[66] L. Mualem,[27] J. Mueller,[43] M. Muether,[216] F. Muheim,[58] A. Muir,[53] M. Mulhearn,[22] D. Munford,[80] L. J. Munteanu,[34] H. Muramatsu,[147] J. Muraz,[52] M. Murphy,[212] T. Murphy,[198] J. Musser,[91] J. Nachtman,[109] Y. Nagai,[60] S. Nagu,[135] M. Nalbandyan,[220] R. Nandakumar,[182] D. Naples,[172] S. Narita,[115] A. Nath,[89] A. Navrer-Agasson,[138] N. Nayak,[19] M. Nebot-Guinot,[58] K. Negishi,[115] A. Nehm,[137] J. K. Nelson,[217] M. Nelson,[109] J. Nesbit,[218] M. Nessi,[66,34] D. Newbold,[182] M. Newcomer,[168] H. Newton,[53] R. Nichol,[209] F. Nicolas-Arnaldos,[73] A. Nikolica,[168] J. Nikolov,[156] E. Niner,[66] K. Nishimura,[79] A. Norman,[66] A. Norrick,[66] P. Novella,[84] J. A. Nowak,[129] M. Oberling,[7] J. P. Ochoa-Ricoux,[23] A. Olivier,[155] A. Olshevskiy,[120] T. Olson,[80] Y. Onel,[109] Y. Onishchuk,[128] A. Oranday,[91] L. Otiniano Ormachea,[45,106] J. Ott,[23] L. Pagani,[22] G. Palacio,[59] O. Palamara,[66] S. Palestini,[34] J. M. Paley,[66] M. Pallavicini,[96,71] C. Palomares,[38] S. Pan,[170] W. Panduro Vazquez,[180] E. Pantic,[22] V. Paolone,[172] V. Papadimitriou,[66] R. Papaleo,[105] A. Papanestis,[182] S. Paramesvaran,[18] A. Paris,[174] S. Parke,[66] E. Parozzi,[98,144] S. Parsa,[13] Z. Parsa,[19] S. Parveen,[118] M. Parvu,[20] D. Pasciuto,[103] S. Pascoli,[57,16] L. Pasqualini,[92,16] J. Pasternak,[88] J. Pater,[138] C. Patrick,[58,209] L. Patrizii,[92] R. B. Patterson,[27] S. J. Patton,[130] T. Patzak,[165] A. Paudel,[66] L. Paulucci,[64] Z. Pavlovic,[66] G. Pawloski,[147] D. Payne,[132] V. Pec,[48] S. J. M. Peeters,[197] A. Pena Perez,[189] E. Pennacchio,[111] A. Penzo,[109] O. L. G. Peres,[29] Y. F. Perez Gonzalez,[57] L. Pérez-Molina,[38] C. Pernas,[217] J. Perry,[58] D. Pershey,[56] G. Pessina,[98] G. Petrillo,[189] C. Petta,[93,30] R. Petti,[190] V. Pia,[92,16] L. Pickering,[180] F. Pietropaolo,[34,101] V. L. Pimentel,[46,29] G. Pinaroli,[19] K. Plows,[160] R. Plunkett,[66] C. Pollack,[174] T. Pollman,[150,3] F. Pompa,[84] X. Pons,[34] N. Poonthottathil,[86] F. Poppi,[92,16] S. Pordes,[66] J. Porter,[197] M. Potekhin,[19] R. Potenza,[93,30] B. V. K. S. Potukuchi,[117] J. Pozimski,[88] M. Pozzato,[92,16] S. Prakash,[29] T. Prakash,[130] C. Pratt,[22] M. Prest,[98] F. Psihas,[66] D. Pugnere,[111] X. Qian,[19] J. L. Raaf,[66] V. Radeka,[19] J. Rademacker,[18] R. Radev,[34] B. Radics,[221] A. Rafique,[7] E. Raguzin,[19] M. Rai,[214] M. Rajaoalisoa,[39] I. Rakhno,[66] L. Rakotondravohitra,[5] R. Rameika,[66] M. A. Ramirez Delgado,[168] B. Ramson,[66] A. Rappoldi,[102,167] G. Raselli,[102,167] P. Ratoff,[129] R. Ray,[66] H. Razafinime,[39] R. F. Razakamiandra,[5] E. M. Rea,[147] J. S. Real,[76] B. Rebel,[218,66] R. Rechenmacher,[66] M. Reggiani-Guzzo,[138] J. Reichenbacher,[191] S. D. Reitzner,[66] H. Rejeb Sfar,[34] A. Renshaw,[80] S. Rescia,[19] F. Resnati,[34] M. Ribas,[199] S. Riboldi,[99] C. Riccio,[194] G. Riccobene,[105] L. C. J. Rice,[172] J. S. Ricol,[76] A. Rigamonti,[34] M. Rigan,[197] E. V. Rincón,[59] A. Ritchie-Yates,[180] S. Ritter,[137] D. Rivera,[133] R. Rivera,[66] A. Robert,[76] J. L. Rocabado Rocha,[84] L. Rochester,[189] M. Roda,[132] P. Rodrigues,[160] M. J. Rodriguez Alonso,[34] J. Rodriguez Rondon,[191] S. Rosauro-Alcaraz,[164] P. Rosier,[164] M. Rossella,[102,167] M. Rossi,[34] M. Ross-Lonergan,[133] J. Rout,[118] P. Roy,[216] C. Rubbia,[74] G. Ruiz Ferreira,[138] B. Russell,[130] D. Ruterbories,[179] A. Rybnikov,[120] A. Saa-Hernandez,[85] R. Saakyan,[209] S. Sacerdoti,[165] S. K. Sahoo,[90] N. Sahu,[90] P. Sala,[99,34] A. R. Samana,[185] N. Samios,[19] O. Samoylov,[120] M. C. Sanchez,[69] P. Sanchez-Lucas,[73] V. Sandberg,[133] D. A. Sanders,[148] D. Sankey,[182] D. Santoro,[99] N. Saoulidou,[10] P. Sapienza,[105] C. Sarasty,[39] I. Sarcevic,[8] I. Sarra,[95] G. Savage,[66] V. Savinov,[172] G. Scanavini,[219] A. Scaramelli,[102] A. Scarff,[188] A. Scarpelli,[19] T. Schefke,[134] H. Schellman,[159,66] S. Schifano,[94,67] P. Schlabach,[66] D. Schmitz,[36] A. W. Schneider,[139] K. Scholberg,[56] A. Schukraft,[66] E. Segreto,[29] A. Selyunin,[120] C. R. Senise,[207] J. Sensenig,[168] M. H. Shaevitz,[44] S. Shafaq,[118] F. Shaker,[221] P. Shanahan,[66] H. R. Sharma,[117] R. Sharma,[19] R. Kumar,[175] K. Shaw,[197] T. Shaw,[66] K. Shchablo,[111] C. Shepherd-Themistocleous,[182] A. Sheshukov,[120] W. Shi,[194] S. Shin,[119] I. Shoemaker,[212] D. Shooltz,[143] R. Shrock,[194] B. Siddi,[94] J. Silber,[130] L. Simard,[164] J. Sinclair,[189] G. Sinev,[191] Jaydip Singh,[135] J. Singh,[135] L. Singh,[47] P. Singh,[176] V. Singh,[47] S. Singh Chauhan,[163] R. Sipos,[34] C. Sironneau,[165] G. Sirri,[92] K. Siyeon,[37] K. Skarpaas,[189] E. Smith,[91] P. Smith,[91] J. Smolik,[49] M. Smy,[23] E. L. Snider,[66] P. Snopok,[87] D. Snowden-Ifft,[157] M. Soares Nunes,[198] H. Sobel,[23] M. Soderberg,[198] S. Sokolov,[120] C. J. Solano Salinas,[106] S. Söldner-Rembold,[138] S. R. Soleti,[130] N. Solomey,[216] V. Solovov,[131] W. E. Sondheim,[133] M. Sorel,[84] A. Sotnikov,[120] J. Soto-Oton,[84] A. Sousa,[39] K. Soustruznik,[35] F. Spagliardi,[160] M. Spanu,[98,144] J. Spitz,[142] N. J. C. Spooner,[188] K. Spurgeon,[198] D. Stalder,[9] M. Stancari,[66] L. Stanco,[101,162] J. Steenis,[22] R. Stein,[18] H. M. Steiner,[130] A. F. Steklain Lisbôa,[199] A. Stepanova,[120] J. Stewart,[19] B. Stillwell,[36] J. Stock,[191] F. Stocker,[34] T. Stokes,[134] M. Strait,[147] T. Strauss,[66] L. Strigari,[201] A. Stuart,[41] J. G. Suarez,[59] J. Subash,[15] A. Surdo,[97] L. Suter,[66] C. M. Sutera,[93,30] K. Sutton,[27] Y. Suvorov,[100,149] R. Svoboda,[22] S. K. Swain,[151] B. Szczerbinska,[202] A. M. Szelc,[58] A. Taffara,[103] N. Talukdar,[190] J. Tamara,[6] H. A. Tanaka,[189] S. Tang,[19] N. Taniuchi,[28] B. Tapia Oregui,[204] A. Tapper,[88] S. Tariq,[66] E. Tarpara,[19] E. Tatar,[82] R. Tayloe,[91] A. M. Teklu,[194] P. Tennessen,[130,4] M. Tenti,[92] K. Terao,[189] F. Terranova,[98,144] G. Testera,[96] T. Thakore,[39] A. Thea,[182] A. Thompson,[201] C. Thorn,[19] S. C. Timm,[66] V. Tishchenko,[19] N. Todorović,[156] L. Tomassetti,[94,67] A. Tonazzo,[165] D. Torbunov,[19] M. Torti,[98,144] M. Tortola,[84] F. Tortorici,[93,30] N. Tosi,[92] D. Totani,[26] M. Toups,[66] C. Touramanis,[132] R. Travaglini,[92] J. Trevor,[27] S. Trilov,[18] W. H. Trzaska,[121] Y. Tsai,[23] Y.-T. Tsai,[189] Z. Tsamalaidze,[72] K. V. Tsang,[189] N. Tsverava,[72] S. Z. Tu,[116] S. Tufanli,[34] C. Tull,[130] J. Turner,[57] M. Tuzi,[84] J. Tyler,[122] E. Tyley,[188] M. Tzanov,[134] M. A. Uchida,[28] J. Urheim,[91] T. Usher,[189] H. Utaegbulam,[198] S. Uzunyan,[153] M. R. Vagins,[123,23] P. Vahle,[217] S. Valder,[197] G. D. A. Valdiviesso,[62] E. Valencia,[77] R. Valentim,[207] Z. Vallari,[27] E. Vallazza,[98] J. W. F. Valle,[84] S. Vallecorsa,[34] R. Van Berg,[168] R. G. Van de Water,[133] D. Vanegas Forero,[141] F. Varanini,[101] D. Vargas Oliva,[205] G. Varner,[79] S. Vasina,[120] N. Vaughan,[159] K. Vaziri,[66] J. Vega,[45] S. Ventura,[101] A. Verdugo,[38] S. Vergani,[28] M. A. Vermeulen,[150] M. Verzocchi,[66] M. Vicenzi,[96,71] H. Vieira de Souza,[165] C. Vignoli,[75] C. Vilela,[34] B. Viren,[19] A. Vizcaya-Hernandez,[43] T. Vrba,[49] Q. Vuong,[179] A. V. Waldron,[176] M. Wallbank,[39] J. Walsh,[143] T. Walton,[66] H. Wang,[24] J. Wang,[191] L. Wang,[130] M. H. L. S. Wang,[66] X. Wang,[66] Y. Wang,[24] K. Warburton,[110] D. Warner,[43] M. O. Wascko,[88] D. Waters,[209] A. Watson,[15]

K. Wawrowska,[182,197] P. Weatherly,[55] A. Weber,[137,66] M. Weber,[13] H. Wei,[134] A. Weinstein,[110] D. Wenman,[218] M. Wetstein,[110] J. Whilhelmi,[219] A. White,[203] A. White,[219] L. H. Whitehead,[28] D. Whittington,[198] M. J. Wilking,[194] A. Wilkinson,[209] C. Wilkinson,[130] Z. Williams,[203] F. Wilson,[182] R. J. Wilson,[43] W. Wisniewski,[189] J. Wolcott,[206] J. Wolfs,[179] T. Wongjirad,[206] A. Wood,[80] K. Wood,[130] E. Worcester,[19] M. Worcester,[19] M. Wospakrik,[66] K. Wresilo,[28] C. Wret,[179] S. Wu,[147] W. Wu,[66] W. Wu,[23] M. Wurm,[137] J. Wyenberg,[54] Y. Xiao,[23] I. Xiotidis,[88] B. Yaeggy,[39] N. Yahlali,[84] E. Yandel,[26] G. Yang,[194] K. Yang,[160] T. Yang,[66] A. Yankelevich,[23] N. Yershov,[107] K. Yonehara,[66] Y. S. Yoon,[37] T. Young,[152] B. Yu,[19] H. Yu,[19] H. Yu,[195] J. Yu,[203] Y. Yu,[87] W. Yuan,[58] R. Zaki,[221] J. Zalesak,[48] L. Zambelli,[52] B. Zamorano,[73] A. Zani,[99] L. Zazueta,[217] G. P. Zeller,[66] J. Zennamo,[66] K. Zeug,[218] C. Zhang,[19] S. Zhang,[91] Y. Zhang,[172] M. Zhao,[19] E. Zhivun,[19] E. D. Zimmerman,[42] S. Zucchelli,[92,16] J. Zuklin,[48] V. Zutshi,[153] and R. Zwaska[66]

(The DUNE Collaboration)

[1]*Abilene Christian University, Abilene, Texas 79601, USA*
[2]*University of Albany, SUNY, Albany, New York 12222, USA*
[3]*University of Amsterdam, NL-1098 XG Amsterdam, The Netherlands*
[4]*Antalya Bilim University, 07190 Döşemealtı/Antalya, Turkey*
[5]*University of Antananarivo, Antananarivo 101, Madagascar*
[6]*Universidad Antonio Nariño, Bogotá, Colombia*
[7]*Argonne National Laboratory, Argonne, Illinois 60439, USA*
[8]*University of Arizona, Tucson, Arizona 85721, USA*
[9]*Universidad Nacional de Asunción, San Lorenzo, Paraguay*
[10]*University of Athens, Zografou GR 157 84, Greece*
[11]*Universidad del Atlántico, Barranquilla, Atlántico, Colombia*
[12]*Augustana University, Sioux Falls, South Dakota 57197, USA*
[13]*University of Bern, CH-3012 Bern, Switzerland*
[14]*Beykent University, Istanbul, Turkey*
[15]*University of Birmingham, Birmingham B15 2TT, United Kingdom*
[16]*Università del Bologna, 40127 Bologna, Italy*
[17]*Boston University, Boston, Massachusetts 02215, USA*
[18]*University of Bristol, Bristol BS8 1TL, United Kingdom*
[19]*Brookhaven National Laboratory, Upton, New York 11973, USA*
[20]*University of Bucharest, Bucharest, Romania*
[21]*University of California Berkeley, Berkeley, California 94720, USA*
[22]*University of California Davis, Davis, California 95616, USA*
[23]*University of California Irvine, Irvine, California 92697, USA*
[24]*University of California Los Angeles, Los Angeles, California 90095, USA*
[25]*University of California Riverside, Riverside California 92521, USA*
[26]*University of California Santa Barbara, Santa Barbara, California 93106 USA*
[27]*California Institute of Technology, Pasadena, California 91125, USA*
[28]*University of Cambridge, Cambridge CB3 0HE, United Kingdom*
[29]*Universidade Estadual de Campinas, Campinas, São Paoulo 13083-970, Brazil*
[30]*Università di Catania, 2–95131 Catania, Italy*
[31]*Universidad Católica del Norte, Antofagasta, Chile*
[32]*Centro Brasileiro de Pesquisas Físicas, Rio de Janeiro, Rio de Janeiro 22290-180, Brazil*
[33]*IRFU, CEA, Université Paris-Saclay, F-91191 Gif-sur-Yvette, France*
[34]*CERN, The European Organization for Nuclear Research, 1211 Meyrin, Switzerland*
[35]*Institute of Particle and Nuclear Physics of the Faculty of Mathematics and Physics of the Charles University, 180 00 Prague 8, Czech Republic*
[36]*University of Chicago, Chicago, Illinois 60637, USA*
[37]*Chung-Ang University, Seoul 06974, South Korea*
[38]*CIEMAT, Centro de Investigaciones Energéticas, Medioambientales y Tecnológicas, E-28040 Madrid, Spain*
[39]*University of Cincinnati, Cincinnati, Ohio 45221, USA*
[40]*Centro de Investigación y de Estudios Avanzados del Instituto Politécnico Nacional (Cinvestav), Mexico City, Mexico*
[41]*Universidad de Colima, Colima, Mexico*
[42]*University of Colorado Boulder, Boulder, Colorado 80309, USA*
[43]*Colorado State University, Fort Collins, Colorado 80523, USA*
[44]*Columbia University, New York, New York 10027, USA*

[45]Comisión Nacional de Investigación y Desarrollo Aeroespacial, Lima, Peru
[46]Centro de Tecnologia da Informacao Renato Archer, Amarais Campinas, São Paulo, CEP 13069-901, Brazil
[47]Central University of South Bihar, Gaya 824236, India
[48]Institute of Physics, Czech Academy of Sciences, 182 00 Prague 8, Czech Republic
[49]Czech Technical University, 115 19 Prague 1, Czech Republic
[50]Dakota State University, Madison, South Dakota 57042, USA
[51]University of Dallas, Irving, Texas 75062-4736, USA
[52]Laboratoire d'Annecy de Physique des Particules, University Grenoble Alpes, University Savoie Mont Blanc, CNRS, LAPP-IN2P3, 74000 Annecy, France
[53]Daresbury Laboratory, Cheshire WA4 4AD, United Kingdom
[54]Dordt University, 700 7th Street NE, Sioux Center, Iowa 51250, USA
[55]Drexel University, Philadelphia, Pennsylvania 19104, USA
[56]Duke University, Durham, North Carolina 27708, USA
[57]Durham University, Durham DH1 3LE, United Kingdom
[58]University of Edinburgh, Edinburgh EH8 9YL, United Kingdom
[59]Universidad EIA, Envigado, Antioquia, Colombia
[60]Eötvös Loránd University, 1053 Budapest, Hungary
[61]Faculdade de Ciências da Universidade de Lisboa—FCUL, 1749-016 Lisboa, Portugal
[62]Universidade Federal de Alfenas, Poços de Caldas, Minas Gerais 37715-400, Brazil
[63]Universidade Federal de Goias, Goiania, Goiás 74690-900, Brazil
[64]Universidade Federal do ABC, Santo André, São Paulo 09210-580, Brazil
[65]Universidade Federal do Rio de Janeiro, Rio de Janeiro 21941-901, Brazil
[66]Fermi National Accelerator Laboratory, Batavia, Illinois 60510, USA
[67]University of Ferrara, Ferrara, Italy
[68]University of Florida, Gainesville, Florida 32611-8440, USA
[69]Florida State University, Tallahassee, Florida, USA
[70]Fluminense Federal University, 9 Icaraí Niterói, Rio de Janeiro 24220-900, Brazil
[71]Università degli Studi di Genova, Genova, Italy
[72]Georgian Technical University, Tbilisi, Georgia
[73]University of Granada and CAFPE, 18002 Granada, Spain
[74]Gran Sasso Science Institute, L'Aquila, Italy
[75]Laboratori Nazionali del Gran Sasso, L'Aquila, Italy
[76]University Grenoble Alpes, CNRS, Grenoble INP, LPSC-IN2P3, 38000 Grenoble, France
[77]Universidad de Guanajuato, Guanajuato 37000, Mexico
[78]Harish-Chandra Research Institute, Jhunsi, Allahabad 211 019, India
[79]University of Hawaii, Honolulu, Hawaii 96822, USA
[80]University of Houston, Houston, Texas 77204, USA
[81]University of Hyderabad, Gachibowli, Hyderabad 500 046, India
[82]Idaho State University, Pocatello, Idaho 83209, USA
[83]Institut de Física d'Altes Energies (IFAE)—Barcelona Institute of Science and Technology (BIST), Barcelona, Spain
[84]Instituto de Física Corpuscular, CSIC and Universitat de València, 46980 Paterna, Valencia, Spain
[85]Instituto Galego de Física de Altas Enerxías, University of Santiago de Compostela, Santiago de Compostela 15782, Spain
[86]Indian Institute of Technology Kanpur, Uttar Pradesh 208016, India
[87]Illinois Institute of Technology, Chicago, Illinois 60616, USA
[88]Imperial College of Science Technology and Medicine, London SW7 2BZ, United Kingdom
[89]Indian Institute of Technology Guwahati, Guwahati, 781 039, India
[90]Indian Institute of Technology Hyderabad, Hyderabad, 502285, India
[91]Indiana University, Bloomington, Indiana 47405, USA
[92]Istituto Nazionale di Fisica Nucleare Sezione di Bologna, 40127 Bologna, Italy
[93]Istituto Nazionale di Fisica Nucleare Sezione di Catania, I-95123 Catania, Italy
[94]Istituto Nazionale di Fisica Nucleare Sezione di Ferrara, I-44122 Ferrara, Italy
[95]Istituto Nazionale di Fisica Nucleare Laboratori Nazionali di Frascati, Frascati, Roma, Italy
[96]Istituto Nazionale di Fisica Nucleare Sezione di Genova, 16146 Genova GE, Italy
[97]Istituto Nazionale di Fisica Nucleare Sezione di Lecce, 73100 Lecce, Italy
[98]Istituto Nazionale di Fisica Nucleare Sezione di Milano Bicocca, 3—I-20126 Milano, Italy
[99]Istituto Nazionale di Fisica Nucleare Sezione di Milano, 20133 Milano, Italy
[100]Istituto Nazionale di Fisica Nucleare Sezione di Napoli, I-80126 Napoli, Italy

[101]*Istituto Nazionale di Fisica Nucleare Sezione di Padova, 35131 Padova, Italy*
[102]*Istituto Nazionale di Fisica Nucleare Sezione di Pavia, I-27100 Pavia, Italy*
[103]*Istituto Nazionale di Fisica Nucleare Laboratori Nazionali di Pisa, Pisa, Italy*
[104]*Istituto Nazionale di Fisica Nucleare Sezione di Roma, 00185 Roma, Italy*
[105]*Istituto Nazionale di Fisica Nucleare Laboratori Nazionali del Sud, 95123 Catania, Italy*
[106]*Universidad Nacional de Ingeniería, Lima 25, Peru*
[107]*Institute for Nuclear Research of the Russian Academy of Sciences, Moscow 117312, Russia*
[108]*University of Insubria, Via Ravasi, 2, 21100 Varese, Italy*
[109]*University of Iowa, Iowa City, Iowa 52242, USA*
[110]*Iowa State University, Ames, Iowa 50011, USA*
[111]*Institut de Physique des 2 Infinis de Lyon, 69622 Villeurbanne, France*
[112]*Institute for Research in Fundamental Sciences, Tehran, Iran*
[113]*Instituto Superior Técnico—IST, Universidade de Lisboa, Lisboa, Portugal*
[114]*Instituto Tecnológico de Aeronáutica, São José dos Campos, Brazil*
[115]*Iwate University, Morioka, Iwate 020-8551, Japan*
[116]*Jackson State University, Jackson, Missouri 39217, USA*
[117]*University of Jammu, Jammu 180006, India*
[118]*Jawaharlal Nehru University, New Delhi 110067, India*
[119]*Jeonbuk National University, Jeonrabuk-do 54896, South Korea*
[120]*Joint Institute for Nuclear Research, Dzhelepov Laboratory of Nuclear Problems 6 Joliot-Curie, Dubna, Moscow Region 141980, Russia*
[121]*University of Jyväskylä, FI-40014 Jyväskylä, Finland*
[122]*Kansas State University, Manhattan, Kansas 66506, USA*
[123]*Kavli Institute for the Physics and Mathematics of the Universe, Kashiwa, Chiba 277-8583, Japan*
[124]*High Energy Accelerator Research Organization (KEK), Ibaraki 305-0801, Japan*
[125]*Korea Institute of Science and Technology Information, Daejeon 34141, South Korea*
[126]*K L University, Vaddeswaram, Andhra Pradesh 522502, India*
[127]*National Institute of Technology, Kure College, Hiroshima 737-8506, Japan*
[128]*Taras Shevchenko National University of Kyiv, 01601 Kyiv, Ukraine*
[129]*Lancaster University, Lancaster LA1 4YB, United Kingdom*
[130]*Lawrence Berkeley National Laboratory, Berkeley, California 94720, USA*
[131]*Laboratório de Instrumentação e Física Experimental de Partículas, 1649-003 Lisboa and 3004-516 Coimbra, Portugal*
[132]*University of Liverpool, Liverpool L69 7ZE, United Kingdom*
[133]*Los Alamos National Laboratory, Los Alamos, New Mexico 87545, USA*
[134]*Louisiana State University, Baton Rouge, Louisiana 70803, USA*
[135]*University of Lucknow, Uttar Pradesh 226007, India*
[136]*Madrid Autonoma University and IFT UAM/CSIC, 28049 Madrid, Spain*
[137]*Johannes Gutenberg-Universität Mainz, 55122 Mainz, Germany*
[138]*University of Manchester, Manchester M13 9PL, United Kingdom*
[139]*Massachusetts Institute of Technology, Cambridge, Massachusetts 02139, USA*
[140]*Max-Planck-Institut, Munich 80805, Germany*
[141]*University of Medellín, Medellín 050026, Colombia*
[142]*University of Michigan, Ann Arbor, Michigan 48109, USA*
[143]*Michigan State University, East Lansing, Michigan 48824, USA*
[144]*Università del Milano-Bicocca, 20126 Milano, Italy*
[145]*Università degli Studi di Milano, I-20133 Milano, Italy*
[146]*University of Minnesota Duluth, Duluth, Minnesota 55812, USA*
[147]*University of Minnesota Twin Cities, Minneapolis, Minnesota 55455, USA*
[148]*University of Mississippi, University, Mississippi 38677, USA*
[149]*Università degli Studi di Napoli Federico II, 80138 Napoli, Italy*
[150]*Nikhef National Institute of Subatomic Physics, 1098 XG Amsterdam, Netherlands*
[151]*National Institute of Science Education and Research (NISER), Odisha 752050, India*
[152]*University of North Dakota, Grand Forks, North Dakota 58202-8357, USA*
[153]*Northern Illinois University, DeKalb, Illinois 60115, USA*
[154]*Northwestern University, Evanston, Illinois 60208, USA*
[155]*University of Notre Dame, Notre Dame, Indiana 46556, USA*
[156]*University of Novi Sad, 21102 Novi Sad, Serbia*
[157]*Occidental College, Los Angeles, California 90041, USA*
[158]*Ohio State University, Columbus, Ohio 43210, USA*

[159]Oregon State University, Corvallis, Oregon 97331, USA
[160]University of Oxford, Oxford OX1 3RH, United Kingdom
[161]Pacific Northwest National Laboratory, Richland, Washington 99352, USA
[162]Universtà degli Studi di Padova, I-35131 Padova, Italy
[163]Panjab University, Chandigarh 160014, India
[164]Université Paris-Saclay, CNRS/IN2P3, IJCLab, 91405 Orsay, France
[165]Université Paris Cité, CNRS, Astroparticule et Cosmologie, Paris, France
[166]University of Parma, 43121 Parma, Italy
[167]Università degli Studi di Pavia, 27100 Pavia, Italy
[168]University of Pennsylvania, Philadelphia, Pennsylvania 19104, USA
[169]Pennsylvania State University, University Park, Pennsylvania 16802, USA
[170]Physical Research Laboratory, Ahmedabad 380 009, India
[171]Università di Pisa, I-56127 Pisa, Italy
[172]University of Pittsburgh, Pittsburgh, Pennsylvania 15260, USA
[173]Pontificia Universidad Católica del Perú, Lima, Perú
[174]University of Puerto Rico, Mayaguez 00681, Puerto Rico, USA
[175]Punjab Agricultural University, Ludhiana 141004, India
[176]Queen Mary University of London, London E1 4NS, United Kingdom
[177]Radboud University, NL-6525 AJ Nijmegen, Netherlands
[178]Rice University, Houston, Texas 77005, USA
[179]University of Rochester, Rochester, New York 14627, USA
[180]Royal Holloway College London, London TW20 0EX, United Kingdom
[181]Rutgers University, Piscataway, New Jersey 08854, USA
[182]STFC Rutherford Appleton Laboratory, Didcot OX11 0QX, United Kingdom
[183]Università del Salento, 73100 Lecce, Italy
[184]San Jose State University, San Jose, California 95192-0106, USA
[185]Universidade Estadual de Santa Cruz, CEP 45662-000, Ilheús, Bahia, Brazil
[186]Sapienza University of Rome, 00185 Rome, Italy
[187]Universidad Sergio Arboleda, 11022 Bogotá, Colombia
[188]University of Sheffield, Sheffield S3 7RH, United Kingdom
[189]SLAC National Accelerator Laboratory, Menlo Park, California 94025, USA
[190]University of South Carolina, Columbia, South Carolina 29208, USA
[191]South Dakota School of Mines and Technology, Rapid City, South Dakota 57701, USA
[192]South Dakota State University, Brookings, South Dakota 57007, USA
[193]Southern Methodist University, Dallas, Texas 75275, USA
[194]Stony Brook University, SUNY, Stony Brook, New York 11794, USA
[195]Sun Yat-Sen University, Guangzhou 510275, China
[196]Sanford Underground Research Facility, Lead, South Dakota 57754, USA
[197]University of Sussex, Brighton BN1 9RH, United Kingdom
[198]Syracuse University, Syracuse, New York 13244, USA
[199]Universidade Tecnológica Federal do Paraná, Curitiba, Brazil
[200]Tel Aviv University, Tel Aviv-Yafo, Israel
[201]Texas A&M University, College Station, Texas 77840, USA
[202]Texas A&M University—Corpus Christi, Corpus Christi, Texas 78412, USA
[203]University of Texas at Arlington, Arlington, Texas 76019, USA
[204]University of Texas at Austin, Austin, Texas 78712, USA
[205]University of Toronto, Toronto, Ontario M5S 1A1, Canada
[206]Tufts University, Medford, Massachusetts 02155, USA
[207]Universidade Federal de São Paulo, 09913-030 São Paulo, Brazil
[208]Ulsan National Institute of Science and Technology, Ulsan 689-798, South Korea
[209]University College London, London WC1E 6BT, United Kingdom
[210]Valley City State University, Valley City, North Dakota 58072, USA
[211]Variable Energy Cyclotron Centre, 700 064 West Bengal, India
[212]Virginia Tech, Blacksburg, Virginia 24060, USA
[213]University of Warsaw, 02-093 Warsaw, Poland
[214]University of Warwick, Coventry CV4 7AL, United Kingdom
[215]Wellesley College, Wellesley, Massachusetts 02481, USA
[216]Wichita State University, Wichita, Kansas 67260, USA
[217]William and Mary, Williamsburg, Virginia 23187, USA
[218]University of Wisconsin Madison, Madison, Wisconsin 53706, USA

[219]Yale University, New Haven, Connecticut 06520, USA
[220]Yerevan Institute for Theoretical Physics and Modeling, Yerevan 0036, Armenia
[221]York University, Toronto, Ontario M3J 1P3, Canada

A primary goal of the upcoming Deep Underground Neutrino Experiment (DUNE) is to measure the $\mathcal{O}(10)$ MeV neutrinos produced by a Galactic core-collapse supernova if one should occur during the lifetime of the experiment. The liquid-argon-based detectors planned for DUNE are expected to be uniquely sensitive to the $\nu_e$ component of the supernova flux, enabling a wide variety of physics and astrophysics measurements. A key requirement for a correct interpretation of these measurements is a good understanding of the energy-dependent total cross section $\sigma(E_\nu)$ for charged-current $\nu_e$ absorption on argon. In the context of a simulated extraction of supernova $\nu_e$ spectral parameters from a toy analysis, we investigate the impact of $\sigma(E_\nu)$ modeling uncertainties on DUNE's supernova neutrino physics sensitivity for the first time. We find that the currently large theoretical uncertainties on $\sigma(E_\nu)$ must be substantially reduced before the $\nu_e$ flux parameters can be extracted reliably; in the absence of external constraints, a measurement of the integrated neutrino luminosity with less than 10% bias with DUNE requires $\sigma(E_\nu)$ to be known to about 5%. The neutrino spectral shape parameters can be known to better than 10% for a 20% uncertainty on the cross-section scale, although they will be sensitive to uncertainties on the shape of $\sigma(E_\nu)$. A direct measurement of low-energy $\nu_e$-argon scattering would be invaluable for improving the theoretical precision to the needed level.

## I. INTRODUCTION

A massive star ($M > 8M_\odot$) employs nuclear fusion to sustain itself by first consuming lighter elements such as hydrogen and helium and later consuming heavier elements. In the canonical narrative, at the end of the star's lifetime, the innermost nickel-iron core can no longer undergo nuclear fusion. Gravity causes the core to collapse into a protoneutron star. Neutron degeneracy stalls the collapse; the core rebounds and produces shock waves which propagate outward from the core. Once the shock waves breach the surface of the star, they expel stellar material and leave behind a compact remnant. This process is referred to as a core-collapse supernova.

A core collapse releases 99% of the star's gravitational potential energy via neutrinos in a prompt burst lasting several seconds [1]. While the protoneutron star traps photons and other particles with electromagnetic and strong interactions, neutrinos easily escape because they interact weakly. The neutrino flux is expected to contain interesting signatures related to different phenomena occurring during a core-collapse supernova [2–6], including insight into the explosion mechanism. While the neutrinos detected from SN1987A [7–10] did help to confirm the basic outline of the core-collapse supernova process, they did not provide tight constraints on astrophysical models. Additional neutrino signals from core-collapse supernovae observed in detectors worldwide [11] will provide data to study the mechanism behind the core collapse, as well as information on the properties of neutrinos themselves.

Obtaining a high-statistics measurement of core-collapse supernova neutrinos is among the primary physics goals for the Deep Underground Neutrino Experiment (DUNE). To detect these low-energy neutrinos, DUNE will utilize its far detector (relative to the beam at Fermilab) located 1.5 km underground at the Sanford Underground Research Facility in South Dakota. The DUNE far detector is currently planned to consist of four liquid argon time-projection chambers (LArTPCs) each with a total volume of around seventeen kilotons [12]. These LArTPC detectors will be sensitive to interactions of neutrinos in the few tens of MeV range [13].

Among large neutrino experiments, DUNE will be uniquely sensitive to the $\nu_e$ component of the supernova signal via the charged-current reaction

$$\nu_e + {}^{40}\mathrm{Ar} \to e^- + {}^{40}K^*. \tag{1}$$

The $\nu_e$ component of the supernova neutrino flux is expected to contain unique features which make its future detection with DUNE a valuable scientific opportunity [12].

The neutrinos generated by a core-collapse supernova have much lower energies (few to tens of MeV) than the GeV-scale neutrino beams of interest for DUNE's accelerator-based oscillation physics program. Below 100 MeV, no measurements of charged-current neutrino-argon cross sections are currently available [14], and competing theoretical calculations have significant discrepancies [15]. While the importance of obtaining a precise understanding of neutrino-nucleus scattering at accelerator energies is widely recognized [17–19], and the impact of related uncertainties has been studied in detail by the DUNE collaboration [20], the same cannot yet be said for the tens of MeV regime relevant for supernova neutrino detection. This situation exists despite shared analysis challenges between the two energy scales; in both cases, a reliable cross-section model is needed for neutrino calorimetry, efficiency estimation, and removal of some classes of background events. Theoretical uncertainties on the cross-section model provide an important limitation on the achievable experimental precision.

In this paper, we examine for the first time the impact of cross-section uncertainties on the interpretation of a possible future observation of supernova neutrinos with DUNE. No attempt is made here to be comprehensive in either the uncertainty budget or in the analysis topics considered; for instance, these studies assume that the distance to the core collapse is known precisely. Our aim is instead to explore how variations of the adopted model of the neutrino-argon cross section affect the results of a measurement of simulated data. The present study is restricted to variations of $\sigma(E_\nu)$, the total charged-current cross section as a function of neutrino energy. The studies presented in this paper use simplified assumptions about detector response, but a realistic efficiency for DUNE includes sensitivity to neutrino energies as low as 5 MeV [21]. Although these studies require an assumption about DUNE's expected energy resolution, similar studies performed in Ref. [12] show that the results are not sensitive to the specific choice of energy resolution [22]. Variations to other aspects of the neutrino interaction model, including predictions of exclusive final-state differential distributions and the description of $^{40}\mathrm{K}^*$ nuclear deexcitations, as well as subdominant neutral-current and $\bar{\nu}_e$ charged-current interactions, are left to future work, both for simplicity and because the related uncertainties are difficult to fully quantify at present.

The algorithm used in our measurements to extract supernova $\nu_e$ flux parameters from simulated DUNE data is presented in Sec. II. In Sec. III, we describe three different procedures for varying the $\nu_e - {}^{40}\mathrm{Ar}$ total cross section, and the impact on the simulated measurements is examined for each approach. We discuss the results, their implications for DUNE's future supernova neutrino effort, and prospects for the future in Sec. IV and conclude in Sec. V.

## II. SUPERNOVA PARAMETER FITTING

### A. Pinched-thermal form

A commonly-used representation for the supernova neutrino fluence (i.e., the time integral of the flux) $\Phi$ passing through the Earth is the pinched-thermal form [23,24]:

$$\Phi(E_\nu) = \frac{\varepsilon}{4\pi d^2}\mathcal{N}\left(\frac{E_\nu}{\langle E_\nu\rangle}\right)^\alpha \exp\left[-(\alpha+1)\frac{E_\nu}{\langle E_\nu\rangle}\right], \quad (2)$$

where

$$\mathcal{N} \equiv \frac{(\alpha+1)^{\alpha+1}}{\langle E_\nu\rangle^2\Gamma(\alpha+1)}, \quad (3)$$

is a normalization constant, $\varepsilon$ is the neutrino luminosity, $E_\nu$ is the neutrino energy, $\langle E_\nu\rangle$ is the mean neutrino energy (related to the temperature of the supernova), and $d$ is the distance from the supernova to Earth. The "pinching parameter" $\alpha$ describes the shape of the tails of the neutrino energy distribution.

The expression in Eq. (2) may be used to represent either an instantaneous flux (with dimensions of neutrinos per area per time) or a fluence in a specific time interval (flux integrated over time, with dimensions of neutrinos per area), depending on the units used for $\varepsilon$. In the instantaneous case, the parameters $\langle E_\nu\rangle$ (MeV) and $\alpha$ (dimensionless) are implicitly time dependent, while for the time-integrated case they should be interpreted as average values. The time-integrated spectrum is also well-described by Eq. (2), and the parameters should be interpreted as being applied to the fluence spectrum over the entire burst. For simplicity, we choose to consider only the time-integrated neutrino flux in which $\varepsilon$ may be expressed in ergs. A distance of $d = 10$ kiloparsecs (kpc) is assumed throughout. Different values of the flux parameters describe each neutrino species separately (i.e., the $\nu_e$ parameters are not the same as the $\bar{\nu}_e$ or $\nu_x \equiv \nu_\mu, \nu_\tau, \bar{\nu}_\mu, \bar{\nu}_\tau$ parameters), but only the $\nu_e$ portion of the flux is of interest for the present study given its dominance in the expected supernova signal in DUNE [12]. For the studies in this paper, we assume equipartition between flavors, i.e., $\alpha_{\nu_e} = \alpha_{\bar{\nu}_e} = \alpha_{\nu_x}$ and $\varepsilon_{\nu_e} = \varepsilon_{\bar{\nu}_e} = \varepsilon_{\nu_x}$, and we adopt the hierarchy in Ref. [25] for the mean neutrino energies. The simulated measurements considered here involve an extraction of the $\nu_e$ pinched-thermal flux parameters $\varepsilon$, $\langle E_\nu\rangle$, and $\alpha$ from the reconstructed neutrino energy spectrum expected for DUNE. Figure 1 shows fluences calculated for a pinched-thermal flux.

### B. SNOwGLoBES

Beyond the neutrino-argon cross section, the supernova signal observed by DUNE will also be affected by the supernova flux, the detector response, efficiency, and energy reconstruction. The SuperNova Observatories with General Long-Baseline Experiment Simulator (SNOwGLoBES) software incorporates the effect of detector response factors, including the cross section, into a simulated supernova neutrino signal. This widely used, open-source event-rate calculation tool offers a quick option to model the DUNE far detector response for supernova neutrino signals [29].

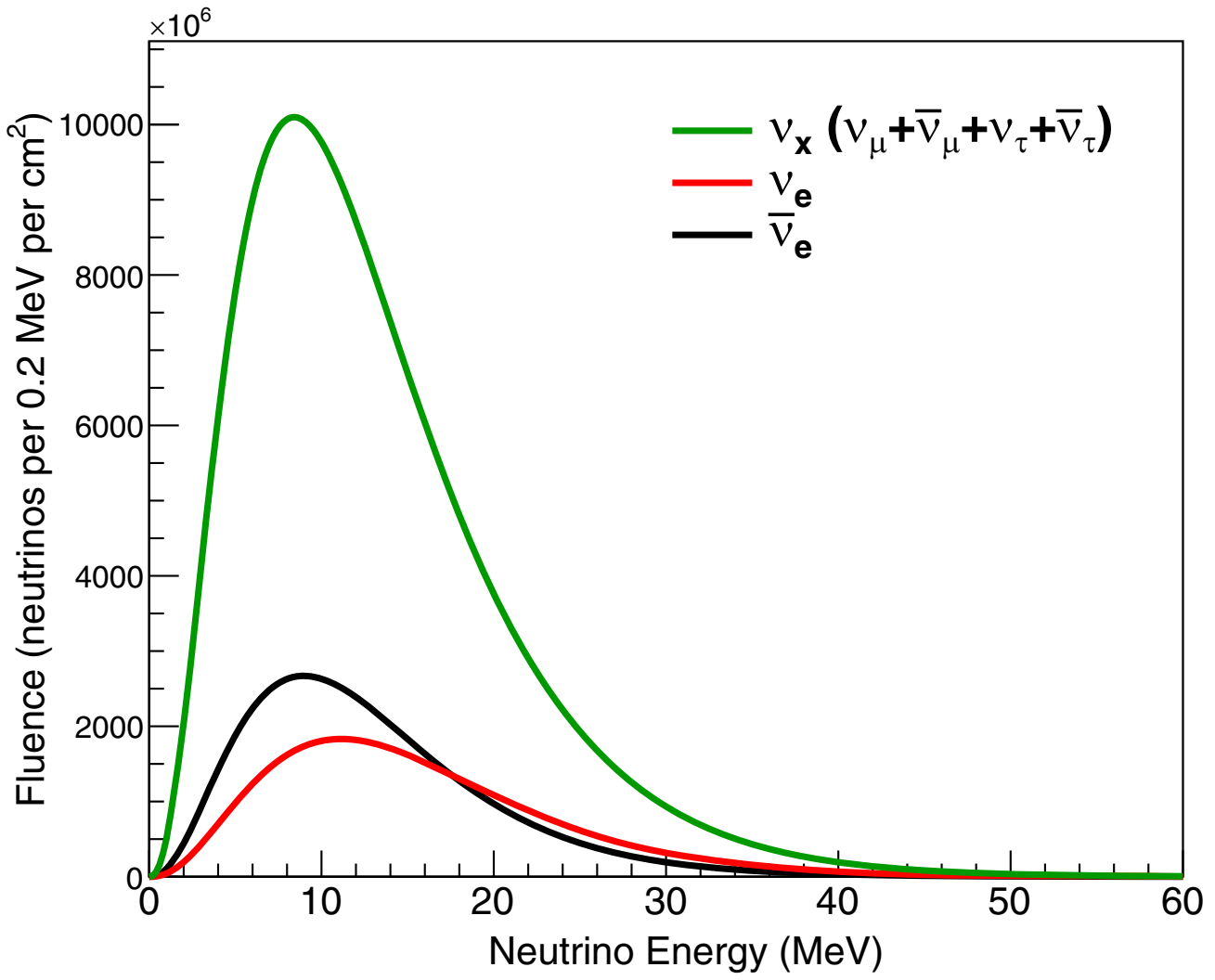

FIG. 1. Pinched-thermal neutrino fluences for a supernova at a distance of 10 kpc. Following Ref. [26], the results are time-integrated over the first ten seconds. The initial fluence parameter values for $\nu_e$ are $(\alpha^0, \langle E_\nu \rangle^0, \varepsilon^0) = (2.5, 9.5\ \mathrm{MeV}, 5 \times 10^{52}\ \mathrm{ergs})$, for $\bar{\nu}_e$ are $(\alpha^0, \langle E_\nu \rangle^0, \varepsilon^0) = (2.5, 12.0\ \mathrm{MeV}, 5 \times 10^{52}\ \mathrm{ergs})$, and for $\nu_x$ are $(\alpha^0, \langle E_\nu \rangle^0, \varepsilon^0) = (2.5, 15.6\ \mathrm{MeV}, 5 \times 10^{52}\ \mathrm{ergs})$. Normal mass ordering and Mikheyev-Smirnov-Wolfenstein (MSW) resonances [27,28] were assumed via Eq. (5).

SNOwGLoBES requires several inputs to perform the simulation, including a cross-section model and a "smearing matrix," i.e., a transfer matrix that can be used to calculate a reconstructed neutrino energy spectrum when applied to the true neutrino energy spectrum (see Fig. 2). In addition, there is an assumed postsmearing detection efficiency. SNOwGLoBES makes use of GLoBES [30] software to convolve a specified flux with a cross section and a smearing matrix. We used fluxes given by Eq. (2) and computed the smearing matrix using simulated $\nu_e - {}^{40}\mathrm{Ar}$ interactions produced by the MARLEY event generator [31,32] with 10% Gaussian smearing applied to the visible energy. The exact value of 10% is modestly optimistic for DUNE's expected capabilities, but the results are not sensitive to the specific value [12].

For our simulated signal predictions, we adopted one of the more optimistic neutrino energy reconstruction scenarios described in Ref. [31]. Under this scenario, the reconstructed neutrino energy is taken to be the *visible energy* $E_{\mathrm{vis}}^{\mathrm{reco}}$ defined by the expression

$$E_{\mathrm{vis}}^{\mathrm{reco}} \equiv E_{\mathrm{bind}}^{\mathrm{min}} + E_{\mathrm{e}} + \mathcal{T}_{\gamma} + \mathcal{T}_{\mathrm{ch}}. \tag{4}$$

Here, $E_{\mathrm{bind}}^{\mathrm{min}} = 0.99$ MeV is the minimum possible change in nuclear binding energy for the charged-current reaction, $E_{\mathrm{e}}$ is the total energy of the outgoing electron, $\mathcal{T}_{\gamma}$ is the summed energy of all de-excitation $\gamma$-rays, and $\mathcal{T}_{\mathrm{ch}}$ is the summed kinetic energy of all final-state charged hadrons. The bimodal behavior of the smearing matrix seen in Fig. 2 is due to neutron emission. Events with final states containing one or more neutrons (assumed to be undetected according to our treatment of $E_{\mathrm{vis}}^{\mathrm{reco}}$) will reconstruct with lower energy.

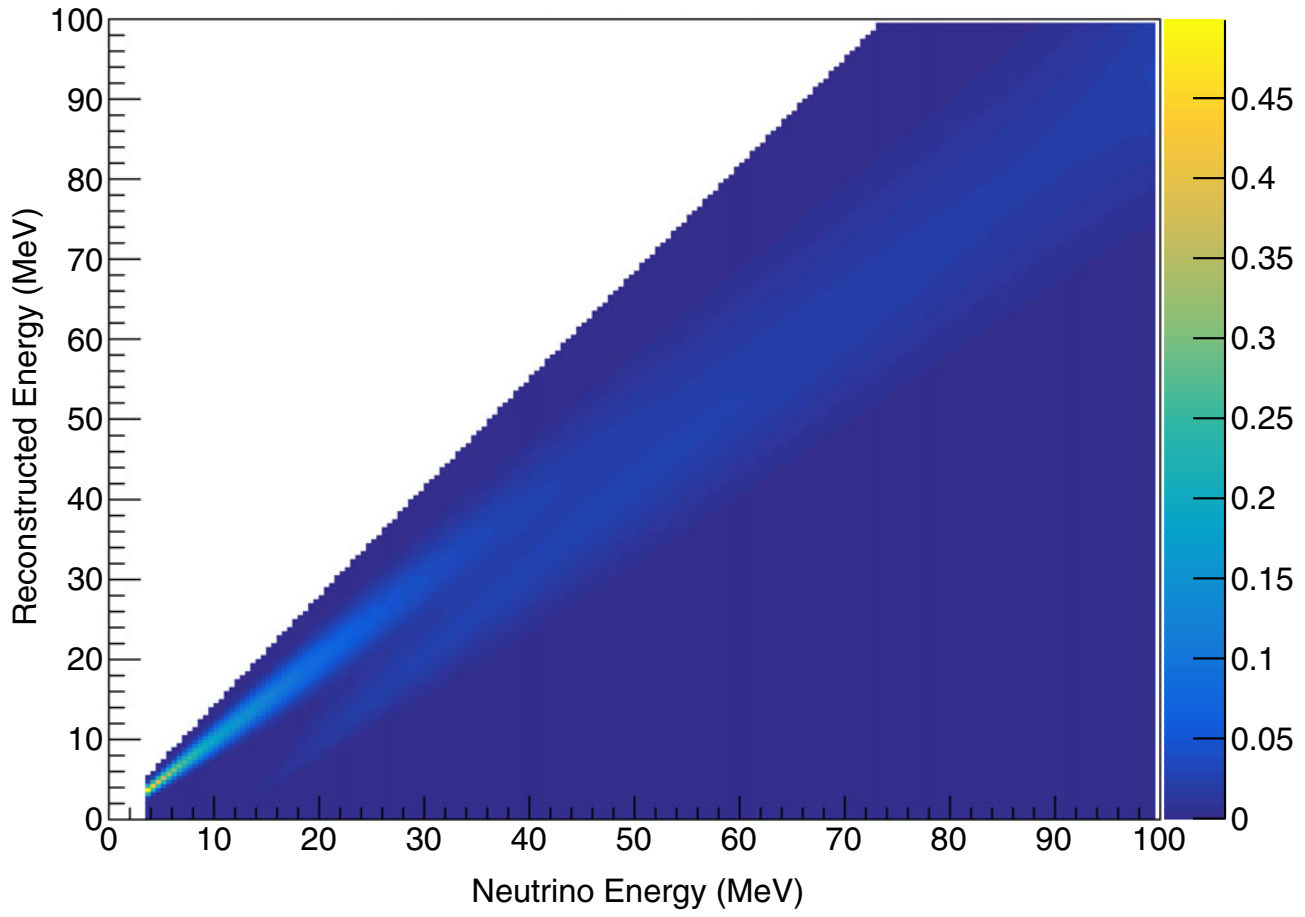

FIG. 2. SNOWGLOBES smearing matrix made with MARLEY modeling and 10% Gaussian-smeared reconstructed energy. An energy column contains the reconstructed energy distribution for neutrino-argon events at a given true neutrino energy.

SNOwGLoBES outputs binned energy spectra (Asimov data sets) corresponding to different detector parameter assumptions and for given pinched-thermal spectral parameters $(\alpha, \langle E_\nu \rangle, \varepsilon)$. Figure 3 shows the two types of SNOwGLoBES output energy spectra; "interaction rates" refers to the energies of neutrinos that interacted (before detector response), while "observed rates" refers to the prediction of the observed spectra in the proposed detector. The observed rates are what the proposed DUNE far detector would observe during the first ten seconds of a 10 kpc supernova burst. The energy loss in the observed rates is due to smearing and neutron emission.

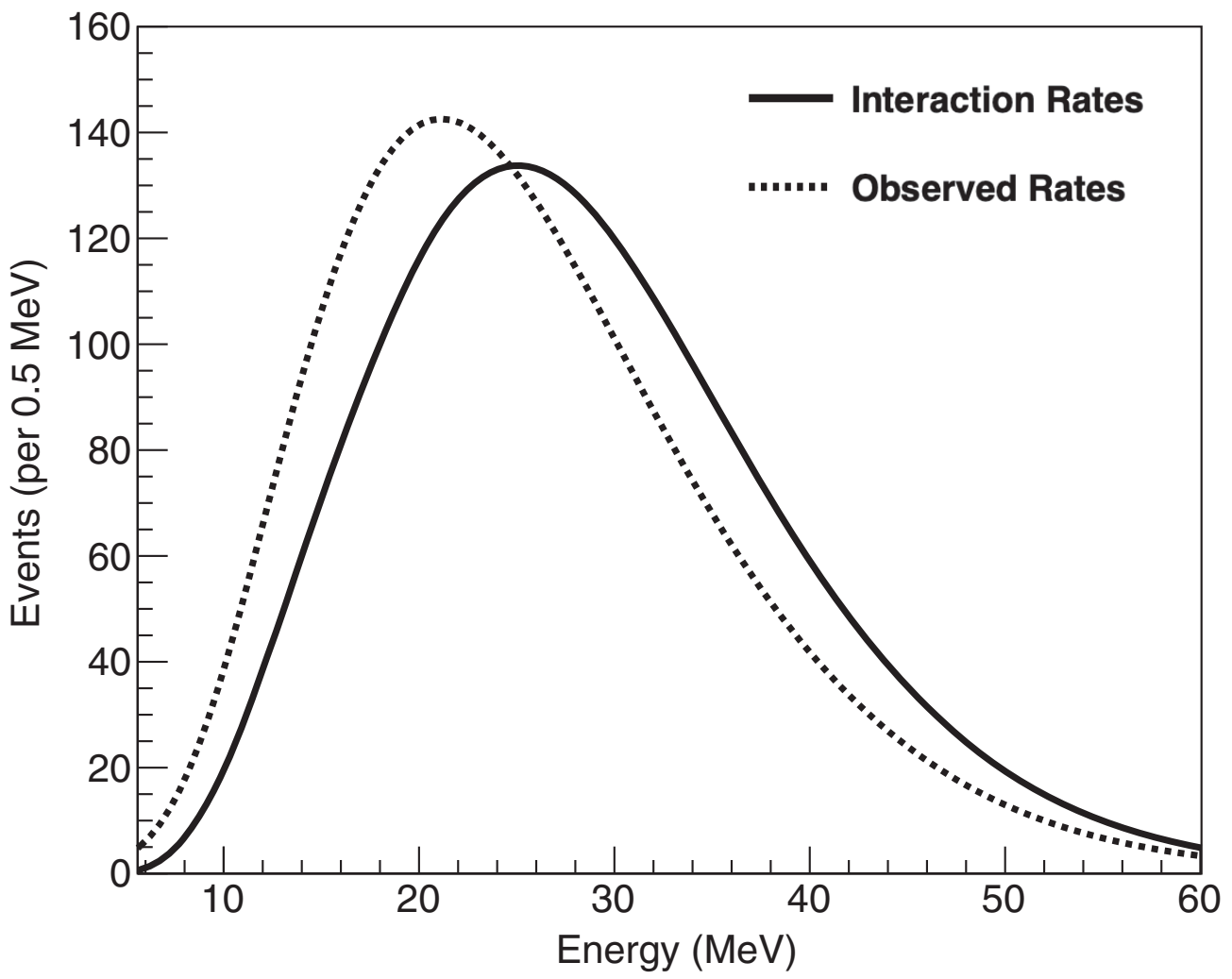

FIG. 3. Interacted and observed event rates calculated using SNOWGLOBES for $\nu_e - {}^{40}\mathrm{Ar}$ interactions in the proposed DUNE far detector. The postsmearing efficiency model imposed a sharp cut at 5 MeV onto the observed rates.

### C. Mass-ordering assumptions in SNOwGLoBES

The different neutrino flavor amplitudes will change as they move through the collapsing star and in the vacuum of space toward Earth. These flavor transitions will affect the $\nu_e$ flux that reaches the DUNE detector, and consequently the flavor transitions will affect the $\nu_e - {}^{40}\mathrm{Ar}$ event rates. SNOwGLoBES provides a simple evaluation of the matter effect for both normal and inverted mass ordering assumptions; we assumed $\theta_{12} = 33.71°$ [33] and the following relations for flavor content for normal mass ordering (NMO) according to the standard prescription in Ref. [34]:

$$F_{\nu_e} = F^0_{\nu_x}, \tag{5a}$$

$$F_{\bar{\nu}_e} = \cos^2(\theta_{12}) F^0_{\bar{\nu}_e} + \sin^2(\theta_{12}) F^0_{\bar{\nu}_x}. \tag{5b}$$

Here, $F_\nu$ is the flux for one (or more) neutrino flavor, and $F^0_\nu$ is the flux before the flavor transition. In the presence of flavor transitions, the observed $\nu_e$ rate at Earth will depend on both the mass ordering and the other produced flavors. To take into account effects produced by flavor transitions, we define a range of flux parameters for $\bar{\nu}_e$ and $\nu_x$ using the $\nu_e$ parameters and the relations outlined in Sec. II A.

### D. Forward fitting

The resulting reconstructed energy spectra from SNOwGLoBES are influenced by the choice of pinched-thermal flux parameters. Measurements of the spectral parameters might contain biases partly introduced by uncertainties in our input assumptions such as the cross-section model. We developed an algorithm that fits a reconstructed neutrino energy spectrum to obtain estimated values of the pinched-thermal parameters; this then enables us to study the impact of the $\nu_e - {}^{40}\mathrm{Ar}$ cross section model on the fit results.

Our algorithm employs a "forward-fitting" approach as an alternative to unfolding; in a forward-fitting approach, a theory prediction convolved with the response of a given detector is compared directly with data. Forward fitting requires two inputs: (1) a reconstructed neutrino energy spectrum produced by SNOwGLoBES for a supernova at a given distance, and (2) a "true" set of pinched-thermal parameters $(\alpha^0, \langle E_\nu \rangle^0, \varepsilon^0)$. The algorithm uses this spectrum as a "true spectrum" to compare against a reference grid of reconstructed energy spectra generated with many different combinations of $(\alpha, \langle E_\nu \rangle, \varepsilon)$. The spectra in the reference grid are also produced by SNOwGLoBES, and the parameter bounds and spacing used in this paper are listed in Sec. II E. In this paper, the true spectrum refers to the assumed true spectrum under test in the algorithm. To quantify goodness-of-fit, the algorithm uses a $\chi^2$ function defined by

$$\chi^2 \equiv \sum_{i=1}^{n_b} \frac{[N_i(\alpha, \langle E_\nu \rangle, \varepsilon) - N_i(\alpha^0, \langle E_\nu \rangle^0, \varepsilon^0)]^2}{\sigma_i^2}. \tag{6}$$

Here $n_b$ is the number of reconstructed energy bins, $N_i$ is the number of events in the $i$th bin, $\sigma_i$ is the statistical uncertainty on the number of events in the $i$th bin of the true spectrum, $(\alpha, \langle E_\nu \rangle, \varepsilon)$ is the set of flux parameters used to generate a reconstructed energy spectrum in the grid, and $(\alpha^0, \langle E_\nu \rangle^0, \varepsilon^0)$ are the flux parameters used to generate the true spectrum. We assume statistics corresponding to the approximately expected flux for a core collapse at 10 kpc.

Figure 4 shows an example comparison of a true spectrum against one arbitrary grid element. Both spectra are represented by Asimov data sets; the error bars of the true spectrum are derived from the Poisson distribution. The true spectrum represents the predicted data that DUNE would observe during a supernova burst.

The collection of $\chi^2$ values for each of the grid elements is used to determine the measurement uncertainty of the pinched-thermal parameters. We consider uncertainty regions in 2D parameter spaces $(\langle E_\nu \rangle, \alpha)$, $(\langle E_\nu \rangle, \varepsilon)$, and $(\alpha, \varepsilon)$. The smallest $\chi^2$ in a given 2D parameter space is determined by profiling over the third parameter, $\varepsilon$, $\alpha$, or $\langle E_\nu \rangle$, respectively. We determine the approximate "sensitivity regions" by placing a cut of $\chi^2 - \chi^2_{\mathrm{min}} = 4.61$ corresponding to a 90% confidence level for two free parameters [33,35]. A sensitivity region is equivalent to the Asimov confidence region for a perfect prediction [36].

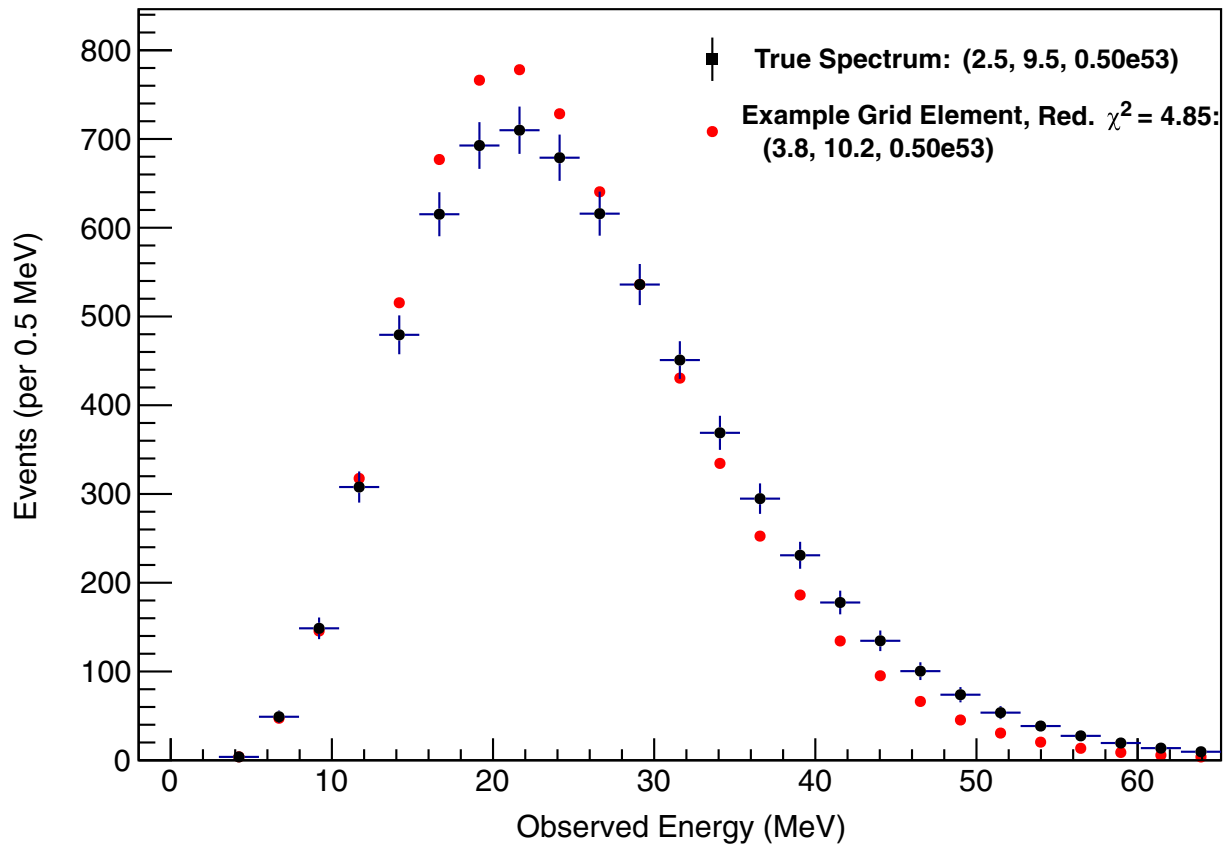

FIG. 4. Event rates calculated using SNOWGLOBES for a true spectrum with initial fluence parameters $(\alpha^0, \langle E_\nu \rangle^0, \varepsilon^0) = (2.5, 9.5\ \mathrm{MeV}, 5 \times 10^{52}\ \mathrm{ergs})$ and an example grid element with fluence parameters $(\alpha^0, \langle E_\nu \rangle^0, \varepsilon^0) = (3.8, 10.2\ \mathrm{MeV}, 5 \times 10^{52}\ \mathrm{ergs})$ and reduced $\chi^2 = 4.85$ based on Eq. (6). The error bars are statistical.

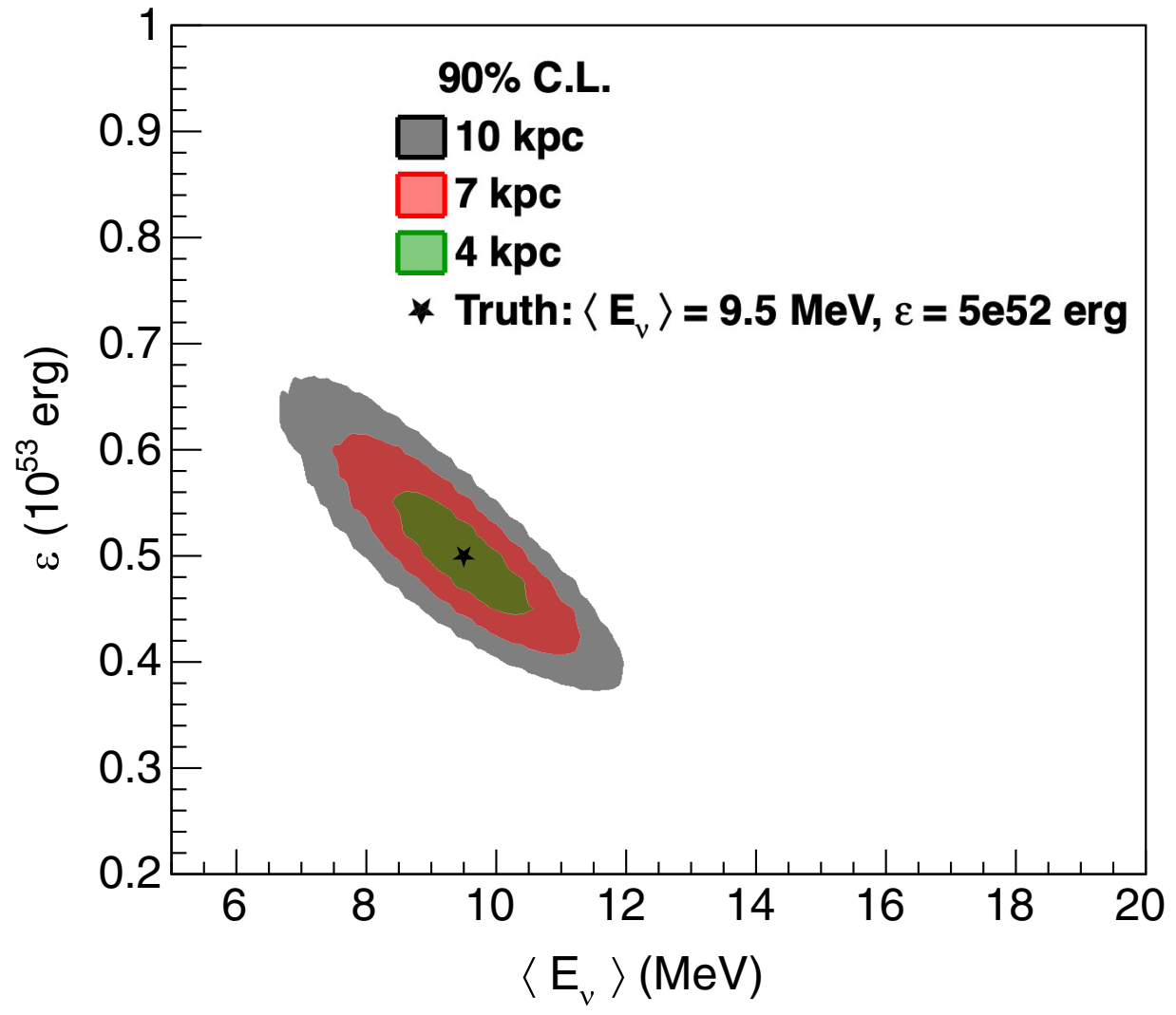

FIG. 5. Sensitivity regions in $(\langle E_\nu\rangle, \varepsilon)$ space for three different supernova distances. These regions were generated from the smearing matrix shown in Fig. 2, a cross section model from MARLEY [32], and a step efficiency function with a 5 MeV detection threshold.

Figure 5 shows sensitivity regions in $(\langle E_\nu\rangle, \varepsilon)$ space for three different supernova distances; the number of events scales with the inverse square of the supernova distance, meaning the regions will grow larger for a more distant supernova.

### E. Figure of merit for forward fitting

We developed a figure of merit as a proxy for the systematic error due to the cross-section uncertainty, where the figure of merit describes the best-fit measurement and characterizes DUNE's expected sensitivity to the supernova flux parameters. The figure of merit $B_x$ is defined as the fractional bias on the measurement of a parameter $x$ obtained from the fitting procedure,

$$B_x \equiv \frac{x^{\mathrm{b.f.}} - x^0}{x^0}. \tag{7}$$

The figure of merit depends on the best-fit value $x^{\mathrm{b.f.}}$ and true value $x^0$ of $x \in \{\alpha, \langle E_\nu\rangle, \varepsilon\}$, where here we express $\langle E_\nu\rangle$ in MeV and $\varepsilon$ in ergs.

For the studies presented in this paper, we define all of our grids using the same range of $\alpha$ and $\langle E_\nu\rangle$ values. The allowed ranges are defined using the $\nu_e$ truth values $(\alpha^0, \langle E_\nu\rangle^0, \varepsilon^0) = (2.5, 9.5, 5\times 10^{52})$ and the following bounds for reasonable $\alpha$ and $\langle E_\nu\rangle$ values are taken from Ref. [26]:

(i) $\alpha \in [0.1, 7.0]$ with 0.1 spacing, corresponding to fractional bias values $B_\alpha \in [-0.96, 1.8]$.

(ii) $\langle E_\nu\rangle \in [5.0, 20.0]$ with 0.1 spacing, corresponding to fractional bias values $B_{\langle E_\nu\rangle} \in [-0.47, 1.10]$.

For the $\varepsilon$ parameter, Ref. [26] defined a reasonable range of $[2\times 10^{52}, 1\times 10^{53}]$ with $2.5\times 10^{51}$ spacing, corresponding to bias values $B_\varepsilon \in [-0.6, 1.0]$. We used this range for the study outlined in Sec. III B. However, for the studies outlined in Secs. III A and III D, this range was insufficient to study the totality of the cross-section space covered by the various $\nu_e - {}^{40}\mathrm{Ar}$ scattering models used in this paper. Therefore, we used the following (more conservative) range of $\varepsilon \in [1.0\times 10^{51}, 1.0\times 10^{54}]$ over several grids with spacings ranging from $2\times 10^{51}$ to $5\times 10^{52}$; the total range of $\varepsilon$ values corresponds to bias values $B_\varepsilon \in [-1.0, 19.0]$.

### F. Study assumptions

Here we summarize the assumptions used for the studies presented in this paper:

(1) All neutrino species contribute to the pinched-thermal flux, where the true parameters for each flavor (before any flavor transition) are defined below [26].
 (a) $\nu_e$ flux: $(\alpha^0, \langle E_\nu\rangle^0, \varepsilon^0) = (2.5, 9.5, 5\times 10^{52})$.
 (b) $\bar{\nu}_e$ flux: $(\alpha^0, \langle E_\nu\rangle^0, \varepsilon^0) = (2.5, 12.0, 5\times 10^{52})$.
 (c) $\nu_x \equiv \nu_\mu, \nu_\tau, \bar{\nu}_\mu, \bar{\nu}_\tau$ flux: $(\alpha^0, \langle E_\nu\rangle^0, \varepsilon^0) = (2.5, 15.6, 5\times 10^{52})$.

(2) A pure pinched-thermal supernova flux.

(3) Normal mass ordering with standard MSW transition effects implemented using Eq. (5); no "collective" effects, spectral swaps, matter effects in the Earth, or nonstandard flavor-transition effects.

(4) A supernova distance of 10 kpc with no distance uncertainty.

(5) Event rates integrated over a supernova burst lasting 10 seconds.

(6) Only charged-current $\nu_e - {}^{40}\mathrm{Ar}$ interactions in the simulated observed signal.

(7) SNOwGLoBES smearing matrix made with MARLEY modeling [32] and 10% Gaussian smearing.

(8) Postsmearing efficiencies in SNOwGLoBES of 100% efficiency above a 5 MeV detection threshold.

### G. Additional information to reproduce the results

The studies in this paper used the following software:

(i) SNOwGLoBES 1.2 [29].

(ii) MARLEY 1.2.0 [32].

(iii) ROOT 6.20 [37].

The studies rely heavily on simulated supernova event rates calculated with SNOwGLoBES. Instructions for how to produce single event rate files, along with grids of flux files, are included in the SNOwGLoBES software package. We used the MARLEY event generator to simulate $\nu_e - {}^{40}\mathrm{Ar}$ interactions while creating a smearing matrix for usage in SNOwGLoBES. The smearing matrix was created using SNOwGLoBES with 10% Gaussian smearing applied. The forward-fitting algorithm and studies were conducted using

ROOT; the forward-fitting algorithm is publicly available on GitHub at https://github.com/erinecon/forward-fitting.

## III. CROSS-SECTION STUDIES

With the forward-fitting algorithm implemented to measure the spectral parameters, construct sensitivity regions, and calculate the bias figure of merit, we studied how the choice of $\nu_e - {}^{40}$Ar cross-section model could impact a supernova neutrino measurement in DUNE. Section III A introduces the various theoretical $\nu_e - {}^{40}$Ar cross section models used in this work. Section III B summarizes a study of one particular cross-section model using a constant overall scaling factor. Section III C details a study over all cross section models available for this work. Finally, the study in Sec. III D considers a restricted range covered by the family of cross-section models. Understanding systematic uncertainties and potential biases introduced by mismodeling of the cross section will be essential for a correct interpretation of any future core-collapse supernova observation.

### A. Neutrino-argon cross-section models

Many calculations of the $\nu_e - {}^{40}$Ar cross section have emerged over time using various nuclear structure models. In the studies performed for this paper, twelve cross-section models are considered. Table I briefly summarizes the features of the models. Figures 6 and 7 show the total charged-current cross sections predicted by each of the models in the energy region of interest. The models were split into two plots for easier readability; the RPA models are all contained in Fig. 6, while the GTBD model and the cross sections calculated by MARLEY are contained in Fig. 7.

The majority of these cross-section models are based on microscopic calculations using formalisms such as the random phase approximation (RPA) or quasiparticle RPA (QRPA). Under these approaches, collective states of nuclei are described using particle-hole (quasiparticle) excitations. The RPA-based calculations include contributions from forbidden (or high-multipole-order) nuclear transitions, which become especially important for neutrinos with $E_\nu > 50$ MeV. A hybrid microscopic calculation [41] in which the allowed (lowest-multipole-order, i.e., Fermi and Gamow-Teller transitions) contributions were computed using the nuclear shell model (NSM) and the forbidden contributions were treated using the RPA is also considered. Alternative macroscopic models like that in Ref. [45] use calculations based on the gross theory of beta decay (GTBD) that describe the global properties of allowed $\beta$-decay processes. The calculations from MARLEY [32] are partially data driven and neglect forbidden nuclear transitions. A QRPA calculation is used by MARLEY at excitation energies where relevant data are not currently available.

TABLE I. Brief features of $\nu_e - {}^{40}$Ar cross-section models used in this work.

| Cross-section model | Model name | Comments |
|---|---|---|
| Default model implemented in SNOWGLOBES [29] | SNOWGLOBES or S | Based on RPA calculations for all multipole transitions up to $J^\pi = 4^\pm$. |
| Calculation by Martinez-Pinedo *et al.* [38,39] | RPA | Based on RPA calculations including all the multipole transitions up to $J^\pi = 6^\pm$. |
| Calculation by Cheoun *et al.* [43] | QRPA-C | Based on QRPA calculations. The results are consistent with data from $(p, n)$ scattering reactions and Gamow-Teller strengths. |
| Calculation by Paar *et al.* [40] | RQRPA | Based on a self-consistent theory framework for a relativistic nuclear energy density functional. The cross sections are including higher-order multipole transitions up to $J^\pi = 5^\pm$. The calculations provide a larger cross sections for $^{40}$Ar. |
| Calculation by Samana *et al.* [42] | PQRPA | Based on projected number QRPA including higher-order multipole transitions up to $J^\pi = 6^\pm$. These calculations were able to describe consistently the weak processes on $^{12}$C [42] using a projection number particle procedure. |
| Calculation by Samana *et al.* [45,46] | GTBD | Based on the gross theory of beta decay, that describes global properties of $\beta$-decay processes. References [45,46] state that this model for heavy elements overestimated available data. Reference [16] states that GTBD is less reliable compared to $(p, n)$ scattering data. |
| Calculation by Suzuki and Honma [41] | NSMRPA or NSM + RPA | Based on a hybrid-model calculation where partial cross sections for Fermi and Gamow-Teller transitions obtained using NSM, while other multipoles computed using RPA calculations. |
| MARLEY calculation based upon $^{40}$Ti $\beta$ decay data [32] | B 1998 | Gamow-Teller matrix elements were extracted from a 1998 measurement by Bhattacharya *et al.* [47]. These are supplemented with QRPA matrix elements from Ref. [43] at high excitation energies. |
| MARLEY calculation based upon an alternative $^{40}$Ti $\beta$ decay data set [32] | L 1998 | Gamow-Teller matrix elements were extracted from a 1998 measurement by Liu *et al.* [48]. These are supplemented with QRPA matrix elements from Ref. [43] at high excitation energies. |
| MARLEY calculation based upon $(p, n)$ scattering data [32] | B 2009 | Gamow-Teller matrix elements were extracted from a 2009 measurement by Bhattacharya *et al.* [49]. These are supplemented with QRPA matrix elements from Ref. [43] at high excitation energies. |
| Unpublished calculation by Samana and dos Santos [44] | QRPA-S | Based on QRPA calculations and using the same parametrization of present PQRPA, including higher-order multipole transitions up to $J^\pi = 6^\pm$. |

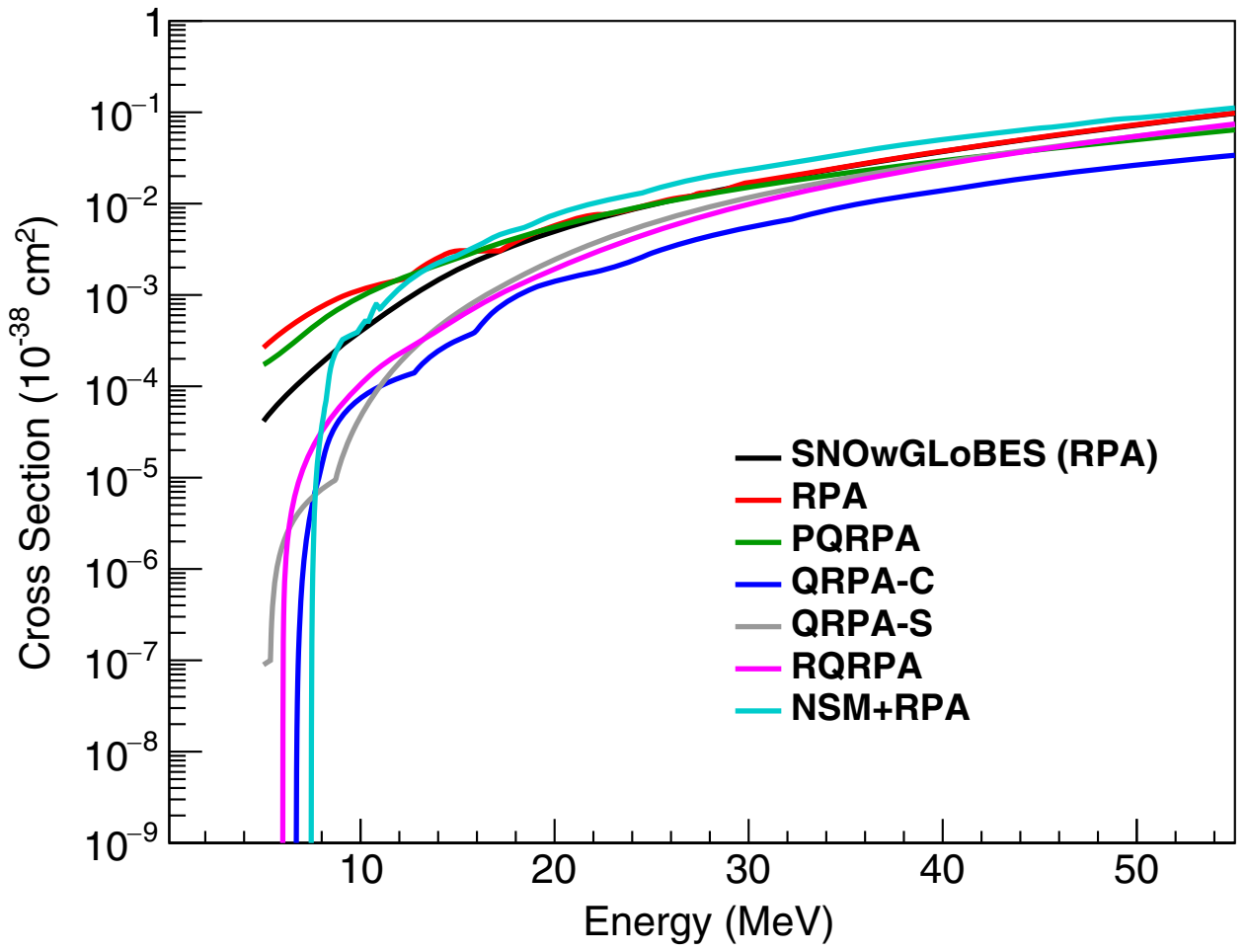

FIG. 6. Cross-section calculations for the $\nu_e - {}^{40}$Ar interaction from Refs. [29,38–44]. The labels are explained in Table I. Note the log scale on the $y$-axis.

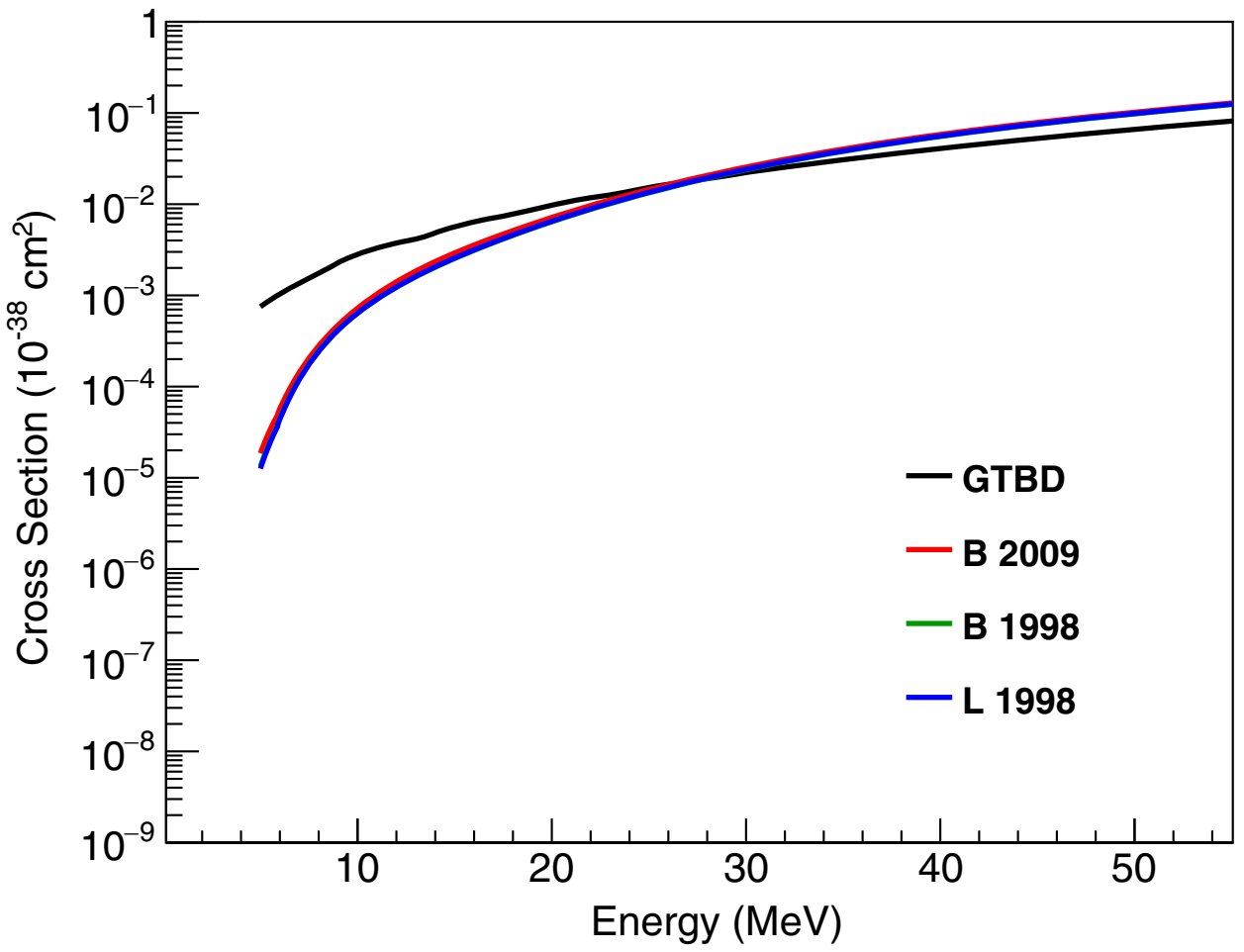

FIG. 7. Cross section calculations for the $\nu_e - {}^{40}$Ar interaction from Refs. [31,45]. The labels are explained in Table I. The $y$-axis range is the same as Fig. 6.

The models include those based on microscopic formalisms such as RPA [38,39], QRPA [43], PQRPA [42], RQRPA [40], and NSM + RPA [41]; macroscopic models such as GTBD [45,46]; and the MARLEY [32] phenomenological calculation based on a Monte Carlo approach. In the absence of any direct measurements of charged-current neutrino-argon scattering in the relevant energy range, experimental constraints on these theoretical approaches are poor. Nevertheless, we can make some general observations about the physics content of these models.

First, all of the microscopic models used here employ different residual interactions. These include the Skyrme interaction (including a spin-orbit term) in the RPA calculation, the Bonn CD potential in QRPA, the $\delta$-interaction in PQRPA, the DDME2 relativistic nuclear-energy density functional in RQRPA, and the monopole-based-universal interaction (VMU) in NSM + RPA. The choice of residual interaction in each case was motivated by a successful description of some relevant experimental data, such as Gamow-Teller (GT) strengths, $\beta$-decay rates, or energies of odd-odd neighboring nuclei.

Second, using a sufficiently large configuration space of nucleon states is important to prevent underestimation of the energy-dependent total cross section $\sigma(E_\nu)$ as the neutrino energy rises. This is due in part to the increasing contribution of higher-order multipoles at high energies. The inclusive or total cross section as function of neutrino energy is a sum over all nuclear multipoles states,

$$\sigma(E_\nu) = \sigma(E_\nu, 0^+) + \sigma(E_\nu, 1^+) + \sigma(E_\nu, 0^-) + \sigma(E_\nu, 1^-) + \sum_{J^\pi \geq 2^\pm}^{J^{\max}} \sigma(E_\nu, J^\pi). \quad (8)$$

Here, $\sigma(E_\nu, J^\pi)$ is the cross section contribution due to multipole $J^\pi$; for example, see Eq. (2.25) in Ref. [50], or Eq. (3) in Ref. [41] for integration over neutrino angle. Usually, the contribution of the multipoles $0^+$ and $1^+$, allowed transitions, are the most important below neutrino energies of 50 MeV. Previous work with PQRPA and RQRPA on $(\nu/\bar{\nu})$ reactions on ${}^{12}$C has examined the variation of $\sigma(E_\nu)$ as a function of the space of single-particle energies and the chosen value of the multipole cutoff $J^{\max}$ [50]. It was found that the magnitudes of the resulting cross sections were close to the sum-rule limit at low energies but significantly smaller than this limit at high energies. As the size of the configuration space is augmented, $\sigma(E_\nu)$ increases steadily, particularly for $(\nu/\bar{\nu})$ energies greater than 200 MeV. Convergence is achieved when the configuration space and multipole cutoff ($J^{\max}$) are both chosen to be sufficiently large [50].

A few words are necessary for the GTBD result. This is a parametric model for $\beta$-decay rates, which includes statistical arguments in a phenomenological way through a convolution between the independent particle model $\beta$-amplitude and the level density of the Fermi gas model corrected to take into account shell effects. The GTBD calculation considers only the contributions of allowed transitions, $\sigma(E_\nu, 0^+)$ and $\sigma(E_\nu, 1^+)$, with a realistic description of the energy of the GT resonance peak [45,46].

Third, some calculations use an effective (or *quenched*) value of the nucleon axial-vector coupling constant for which its bare value $g_A = 1.2756$ from the experimental data [33] is multiplied by a factor of around 0.8. There is still a lack of consensus in the nuclear physics community about whether this quenching is needed. For the family of models considered in this paper, the RPA calculations do not use a renormalization of $g_A$ [39], while the RQRPA model used $g_A = 1$. The PQRPA calculations also adopted $g_A = 1$ to be consistent with comparisons of 2s1d and 2p1f shell- model predictions with measured allowed $\beta$-decay rates [50] and with recent double-beta decay calculations. The QRPA calculations reported in Ref. [43] use a universal quenching factor $f_q = g_A^{\mathrm{eff}}/g_A = 0.74$ to reproduce measured GT strength distributions. The NSM + RPA calculations within the VMU potential used a similar quenching factor $f_q = 0.775$ with $g_A = 1.263$. This choice enabled the NSM + RPA model to describe the experimental cumulative sum of the GT strength rather well. On the other hand, recent studies on variations of $g_A$ in the GTBD have shown that best results for a set of 94 nuclei of interest are obtained with $g_A = 1$ [51]. The GT distribution used for the NSM + RPA calculation is shifted toward higher energy values with significantly smaller strengths for <10 MeV neutrino energies, resulting in a characteristic cutoff at energies below about 8 MeV.

Despite the differences explored above, the main features of measured weak-interaction observables, such as $\beta$-decay strengths and inclusive muon capture rates, are reasonably well described for multiple nuclei by the majority of the nuclear structure models considered herein. By incorporating these cross-section models into our SNOwGLoBES calculations, we studied the impact of variations in the shape of $\sigma(E_\nu)$ on the simulated measurements of supernova neutrino flux parameters. Many of the cross section models required reformatting with extra data points for usage in SNOwGLoBES; Appendix A provides more details on the interpolation procedure that was used. Figures 6 and 7 show that the cross-section models differ considerably and lead to a wide range of predictions for the supernova $\nu_e$ signal in DUNE. Appendix B provides a table of the corresponding event rates as output by SNOwGLoBES (see Table IV).

Figure 8 shows representative expected event rates in DUNE for the CC $\nu_e - {}^{40}\mathrm{Ar}$ absorption process and a supernova at a distance of 10 kpc from Earth. The large differences in the cross-section model predictions at low neutrino energy translate to large variations in the plotted observed energy distributions. Apart from effects of cross-section mismodeling (which are considered in the next section), the expected statistical uncertainty on the event rate has a strong effect on the precision with which the supernova flux parameter values may be measured. The sensitivity regions shown in Fig. 9 are obtained by considering the statistical uncertainty and using the same cross-section model to generate the fake data and extract the results. The GTBD cross section model, which predicts 7770 $\nu_e$CC events, results in the tightest constraints on the flux parameters. The QRPA-C model predicts 1383 events and thus provides the loosest constraints.

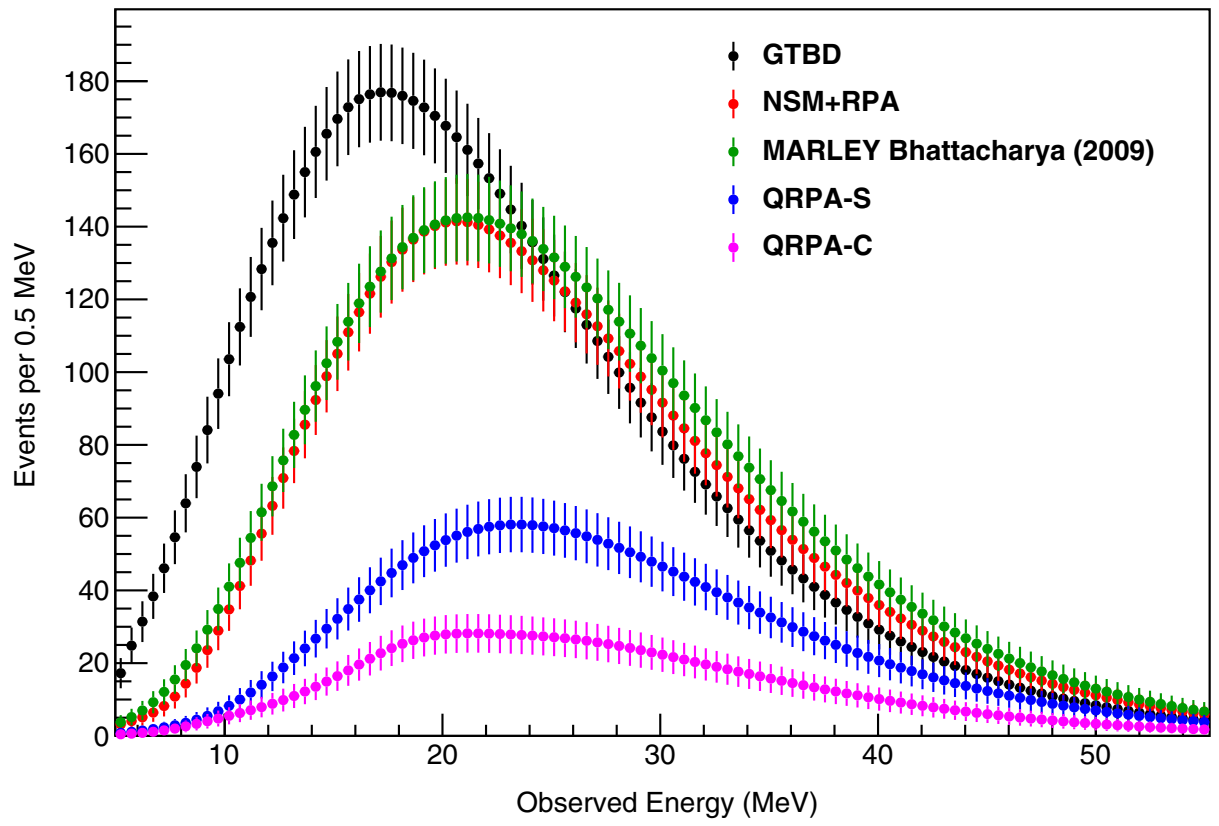

FIG. 8. SNOWGLOBES event rates for select cross-section calculations from Refs. [31,41,43–46]. The initial fluence parameter values for $\nu_e$ are $(\alpha^0, \langle E_\nu\rangle^0, \varepsilon^0) = (2.5, 9.5\ \mathrm{MeV}, 5\times10^{52}\ \mathrm{ergs})$, for $\bar{\nu}_e$ are $(\alpha^0, \langle E_\nu\rangle^0, \varepsilon^0) = (2.5, 12.0\ \mathrm{MeV}, 5\times10^{52}\ \mathrm{ergs})$, and for $\nu_x$ are $(\alpha^0, \langle E_\nu\rangle^0, \varepsilon^0) = (2.5, 15.6\ \mathrm{MeV}, 5\times10^{52}\ \mathrm{ergs})$. Normal mass ordering and MSW resonance were assumed. Note that "QRPA-C" and "QRPA-S" contain the same type of calculation performed by different groups, with the former by Cheoun *et al.* [43] and the latter by Samana and dos Santos [44]. More details about the various models are provided in Table I. The error bars are statistical.

### B. Cross-section normalization uncertainty

As a first examination of the impact of cross-section uncertainties on the extraction of supernova flux parameters from a future DUNE data set, we consider model variations that involve the application of a constant overall scaling factor. These variations shift a plot of $\sigma(E_\nu)$ vertically while leaving the shape unchanged (see Fig. 10). We adopt as a reference model a cross section from MARLEY version 1.2.0 [31,52].

The data-driven nuclear matrix elements in this model were obtained from a measurement of very forward $(p, n)$ scattering reported in Ref. [49]. The unaltered reference model is used together with versions changed by factors of $\pm(5\ \mathrm{to}\ 20)\%$ in 5% steps, $\pm50\%$, and $+100\%$. This procedure yields a total of twelve unique cross-section models, and those models generate different true spectra and grids that we used as input into the forward-fitting algorithm.

Figure 11 shows sensitivity regions for a 10 kpc supernova, the true scenario outlined in Sec. II F, and three different sets of assumptions. The sensitivity regions shift for changes in $\varepsilon$; the cross-section scaling factors affect the statistics and thus $\varepsilon$. The sensitivity regions shift vertically for change in cross-section normalization, with near-negligible shape change, as expected.

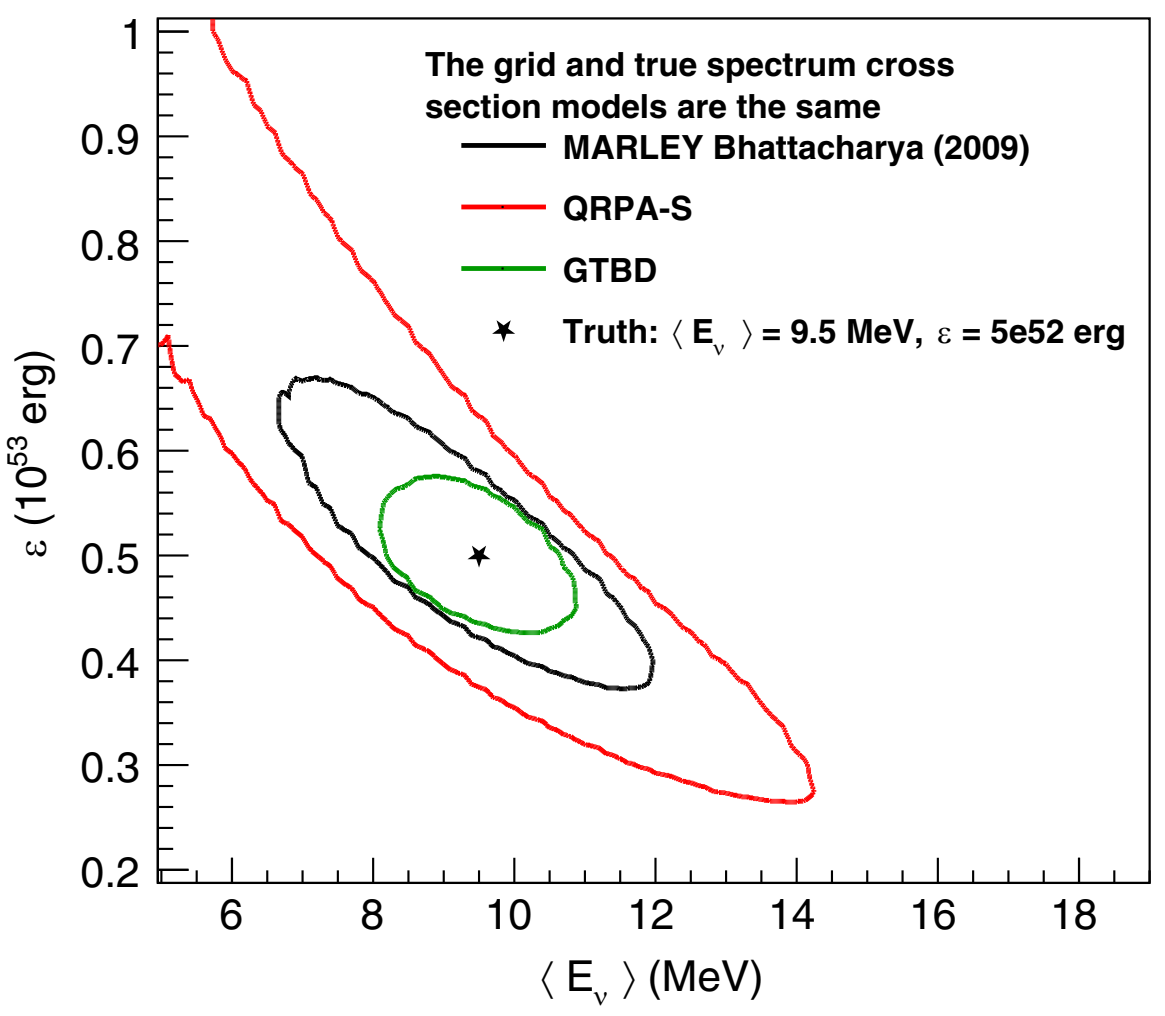

FIG. 9. Sensitivity regions (90% C.L.) in $(\langle E_\nu \rangle, \varepsilon)$ space generated from the cross-section models in Refs. [31,44,45]. Only statistical uncertainties are considered. In each case, the same cross-section model is used both to produce the fake data and to calculate the sensitivity region.

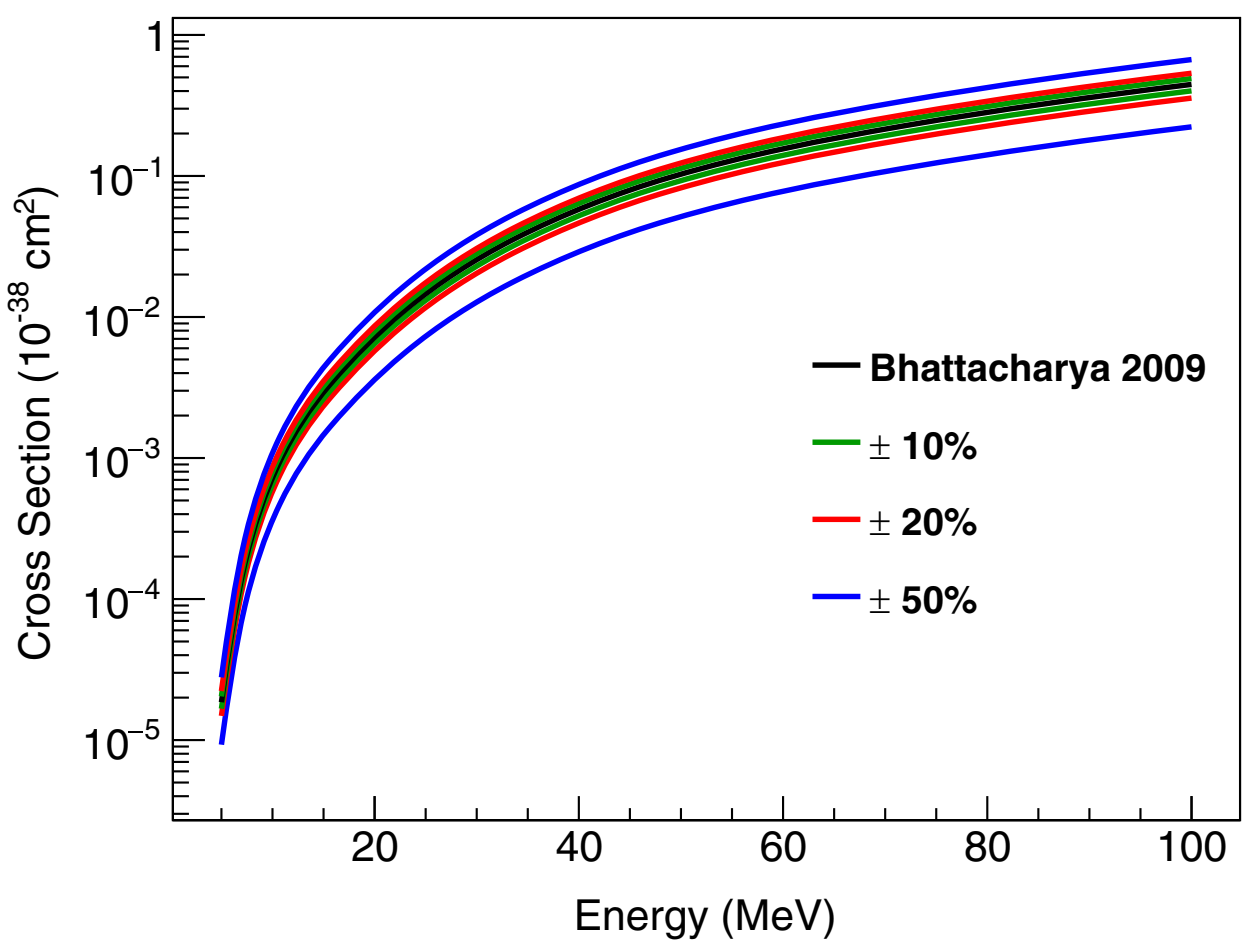

FIG. 10. $\nu_e - {}^{40}$Ar cross section versus energy with various scaling factors applied. Reference [31] provided the cross-section model using nuclear matrix element data from Ref. [49].

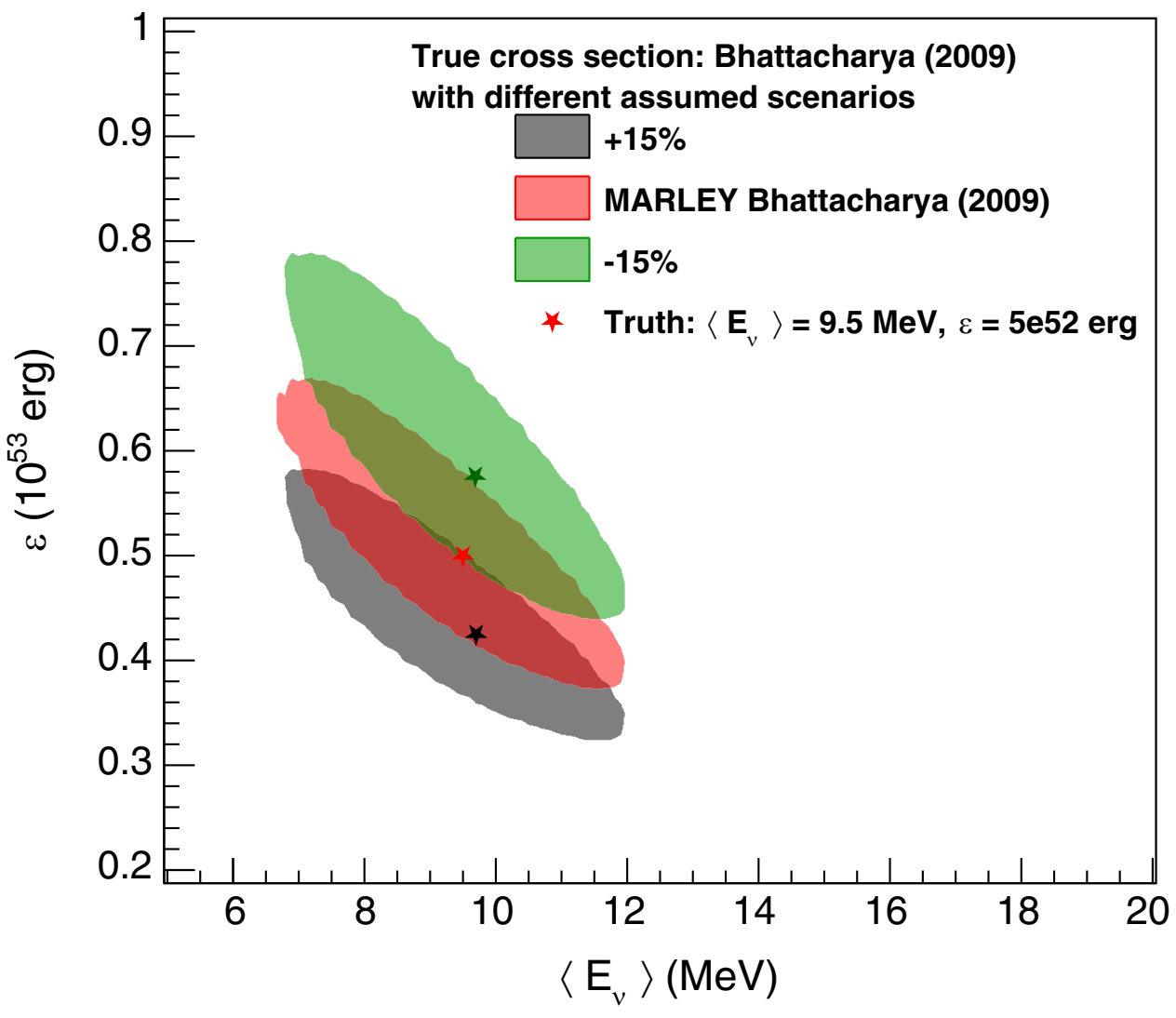

FIG. 11. Sensitivity regions (90% C.L.) for a 10 kpc supernova to study different combinations of assumed and true total cross section normalizations.

Figure 12 shows the bias in the best-fit parameter values for each possible combination of true cross-section model (i.e., the model used to simulate the fake dataset) and assumed cross-section model (i.e., the model used to perform the parameter fits). The best fit within the grid bounds is determined, and that constraint can introduce an artificial bias to the best fit once a boundary is reached for one or more parameters. The results are shown separately for $\alpha$, $\langle E_\nu \rangle$, and $\varepsilon$. For each parameter, a two-dimensional histogram is plotted in which each bin represents a particular combination of cross-section models. The color of the bin represents the bias value, i.e., the fractional difference between the best-fit parameter value and its true value.

We first notice that the biases on $\alpha$ and $\langle E_\nu \rangle$ are relatively small unless the assumptions significantly differ from reality. If we assume an enhanced cross section (using positive scaling factors), the large mismatch in statistics causes an $\varepsilon$ underestimation. The difference in statistics forces the algorithm to select lower $\varepsilon$ values. If we assume a reduced cross section (using negative scaling factors), we expect a lower event rate than we actually observe; thus the forward-fitting algorithm prefers higher $\varepsilon$ values to compensate for the discrepancy. When the algorithm reaches a boundary (i.e., at the minimum or maximum $\varepsilon$ value allowed), the biases in $\alpha$ and $\langle E_\nu \rangle$ will increase to compensate for spectral shape differences between the true spectrum and grid elements.

### C. Combined cross-section normalization and shape uncertainty

To characterize the impact of using an inaccurate cross-section model to extract values of the supernova flux parameters, we consider scenarios in which different combinations of the theoretical models described in Sec. III A are used to (1) simulate a fake data set, and (2) perform fits of the flux parameters. Figure 13 displays the 2D bias plots for the different combinations of assumed and true total cross-section models. A logarithmic color scale is used for $\varepsilon$ due to the very large range of biases allowed for that parameter. In the 2D plots, the cross-section models are ordered along each histogram axis from smallest to largest expected number of events integrated over a neutrino-energy range of [5, 15] MeV. Appendix B also contains the numerical values for the expected event counts for each model in the [5, 15] MeV range.

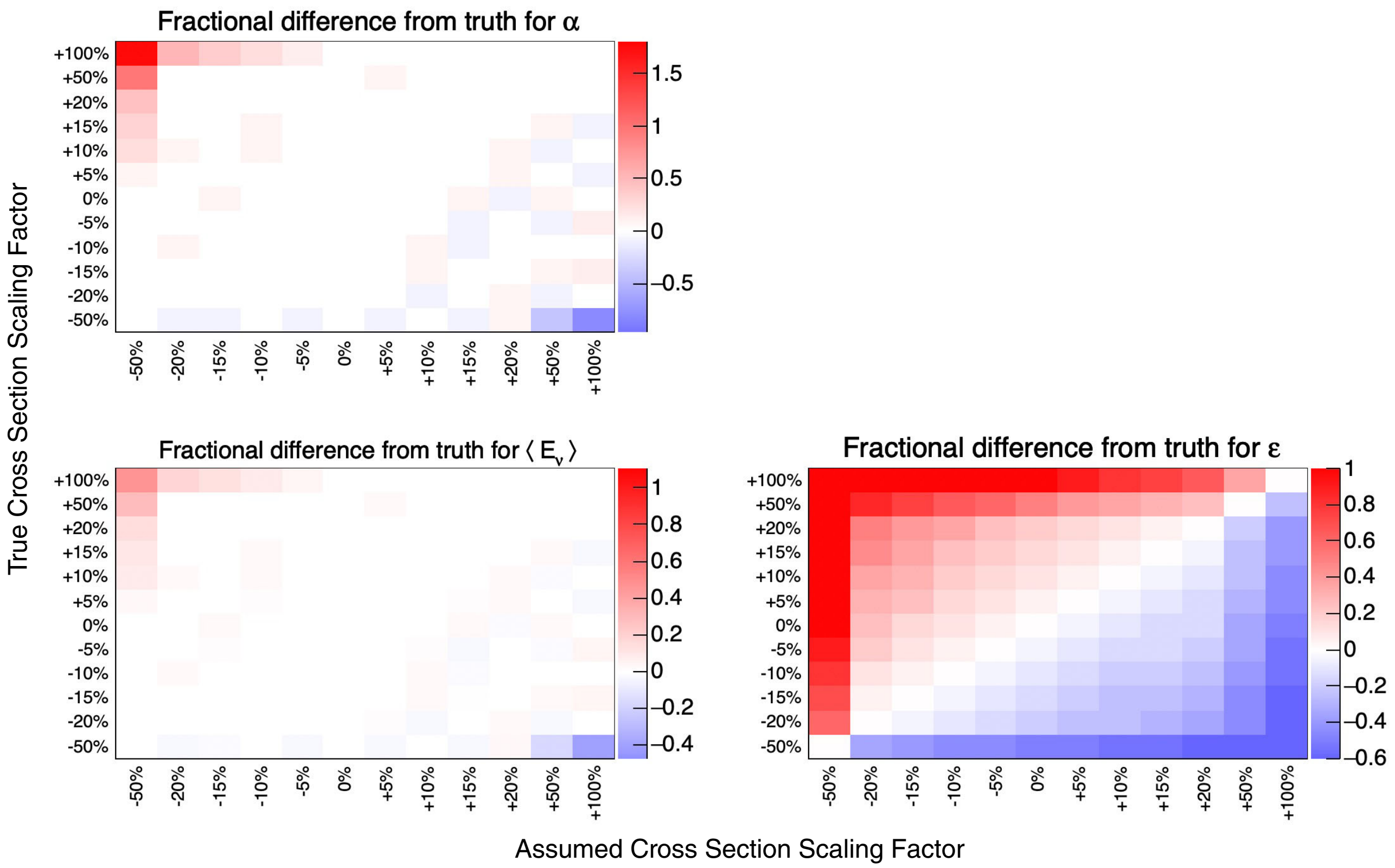

FIG. 12. 2D fractional difference plots to study effects produced by normalization uncertainties on the total cross section.

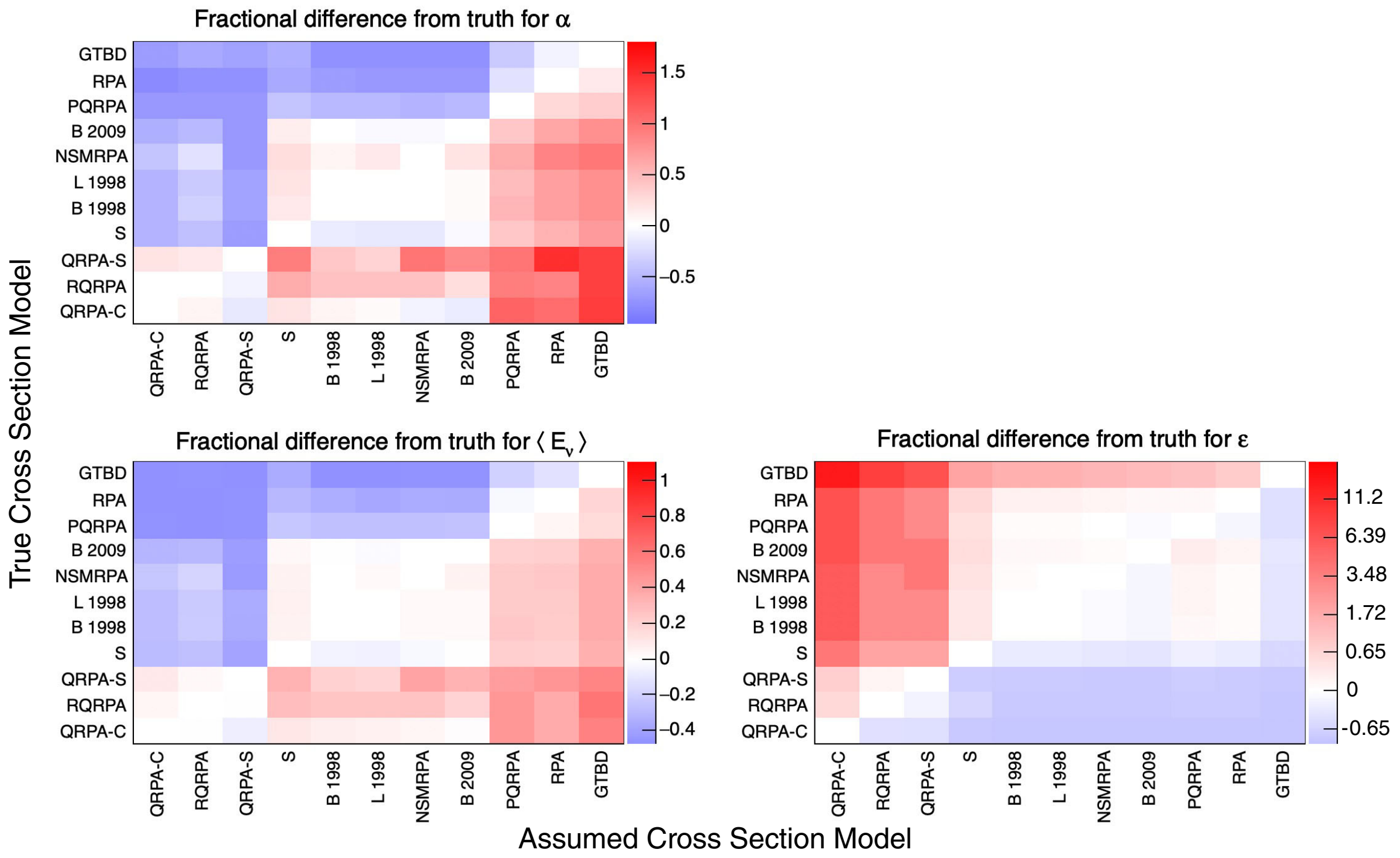

FIG. 13. 2D fractional difference plots to study effects produced by different cross section models. Note that “S” stands for the cross section model implemented into SNOWGLOBES [29]. Also note that the $\varepsilon$ color-scale is log to account for the wide range of values. The number scale shows the raw fractional difference values to conform with the $\alpha$ and $\langle E_\nu \rangle$ plots.

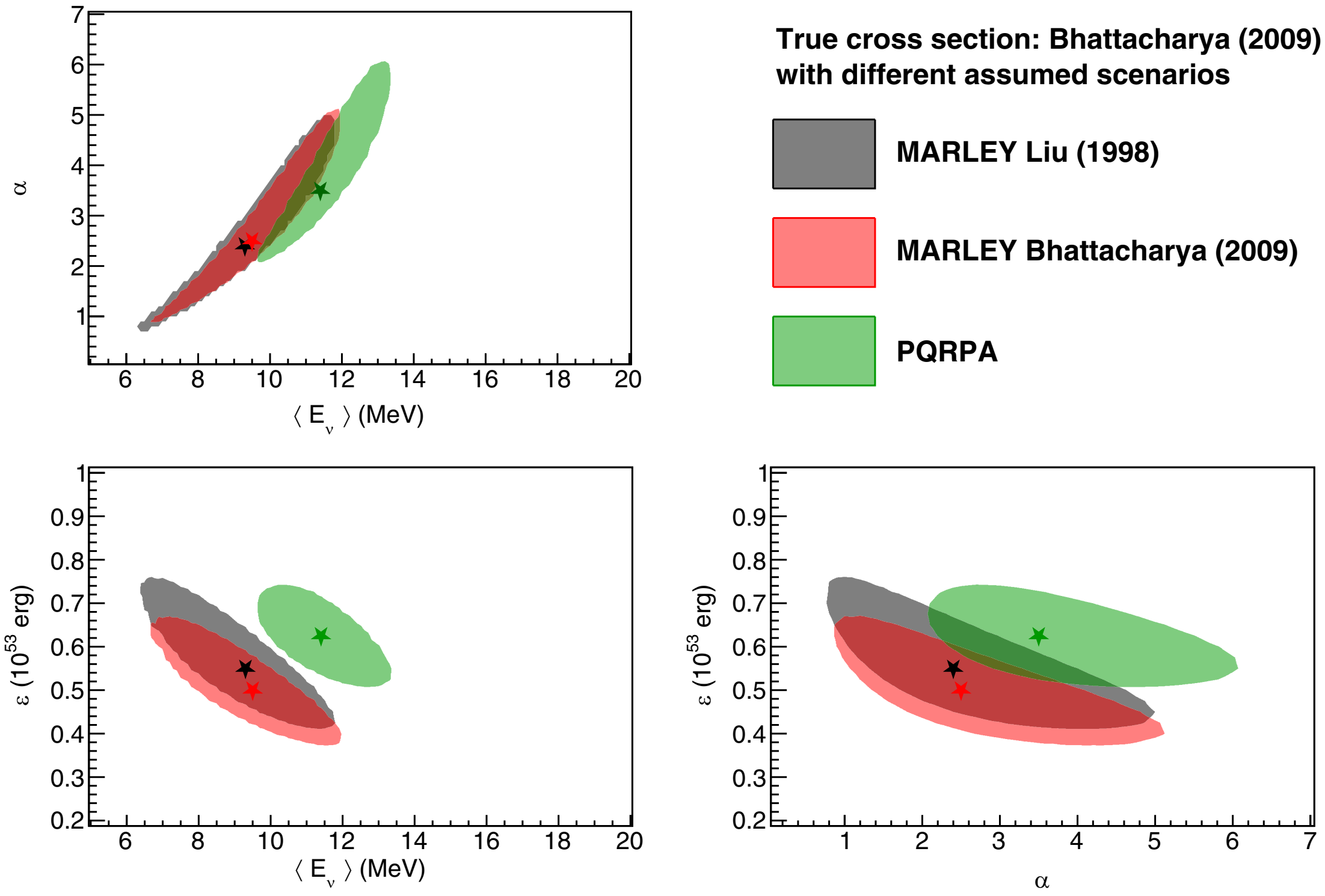

FIG. 14. Sensitivity regions (90% C.L.) calculated with different assumed cross-section models for a fake data set generated using the MARLEY B 2009 model. The stars mark the best-fit measurements from the fitting algorithm. The red stars also indicate the true parameter values, i.e., when the assumed cross section model is identical to the true model.

Further insight into cross-section model effects on the extraction of supernova neutrino flux parameters can be gained from Fig. 14, which shows sensitivity regions computed based on a fake dataset produced using the MARLEY B 2009 cross-section model. When supernova flux parameters are extracted using the same cross-section model (red sensitivity regions), the best-fit values (red stars) are identical to the true ones by construction. A small bias is seen when the extraction procedure is repeated using the MARLEY L 1998 model (black stars). However, the difference between the assumed (L 1998) and true (B 2009) cross sections is small enough that the gray sensitivity regions obtained from the new fit cover the true parameter values in all cases. A more problematic bias (green stars) is seen when the fit is repeated using the PQRPA model as the assumed cross section. In this case, the difference between the PRQPA and MARLEY B 2009 predictions is large enough to lead to green sensitivity regions which do not enclose the true results. This bias would need to be corrected in the context of a real analysis by introducing a cross-section-related systematic uncertainty to inflate the sensitivity regions. The significant corresponding loss of precision can be visually estimated from Fig. 14 by examining the degree to which the green sensitivity regions "miss" the red star that represents the true parameter values.

Some general trends were seen in the course of these fake data studies. If the cross-section model used for fitting gives higher values than the true one used to generate the fake data, then the fitting algorithm tends to overestimate $\alpha$ and $\langle E_\nu \rangle$ while underestimating $\varepsilon$. Because $\varepsilon$ is directly proportional to the expected number of events, the best-fit value of $\varepsilon$ is driven lower for fake data sets with low statistics.

### D. Total cross-section uncertainty envelope

The cross-section models considered above are not expected to produce results of equal quality in the energy region of interest for supernova neutrinos (see, e.g., the discussion in the supplemental materials from Ref. [16]), and furthermore, uncertainties are typically not available for them. As a means of assigning a theoretical uncertainty which neglects implausibly extreme variations, we consider the spread between three cross-section predictions; the partially data-driven MARLEY models [31], the NSM + RPA calculation [41], and the QRPA-S calculation [44]. In the absence of a direct measurement of the $\nu_e$ capture process on argon, we selected this subset of the available models based upon purely *a priori* considerations. Predictions from our chosen subset of cross-section models are shown in Fig. 15. An uncertainty envelope defined as the range between the minimum and maximum cross-section predictions from this subset of models is also shown as the crosshatched region. Predicted supernova neutrino event rates in DUNE for each of the models used to define the envelope are displayed in Fig. 16.

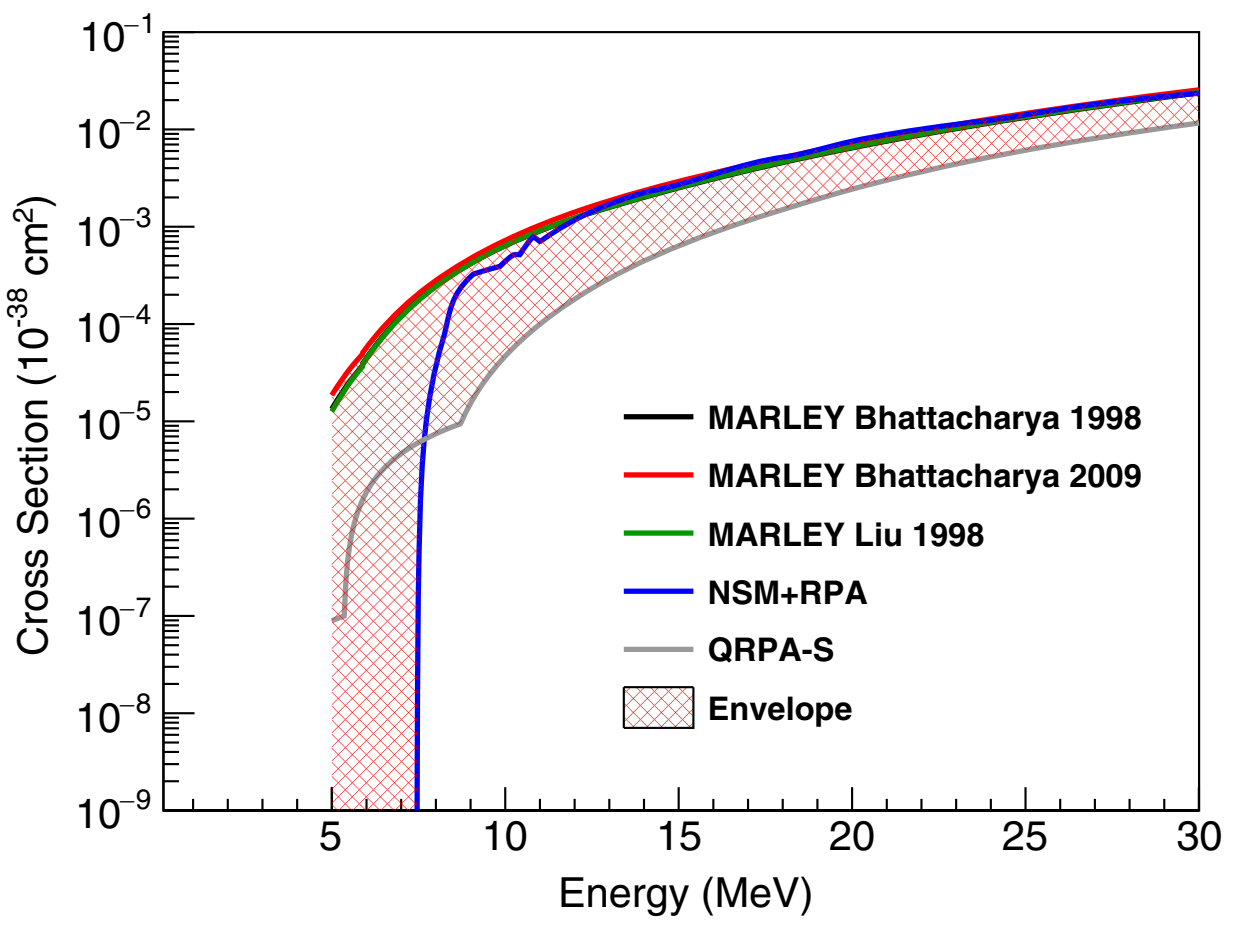

FIG. 15. Total cross section predictions for the $\nu_e - {}^{40}$Ar interaction from the selected subset of models discussed in Sec. III D. The shaded region represents the adopted uncertainty envelope based on the spread of these models.

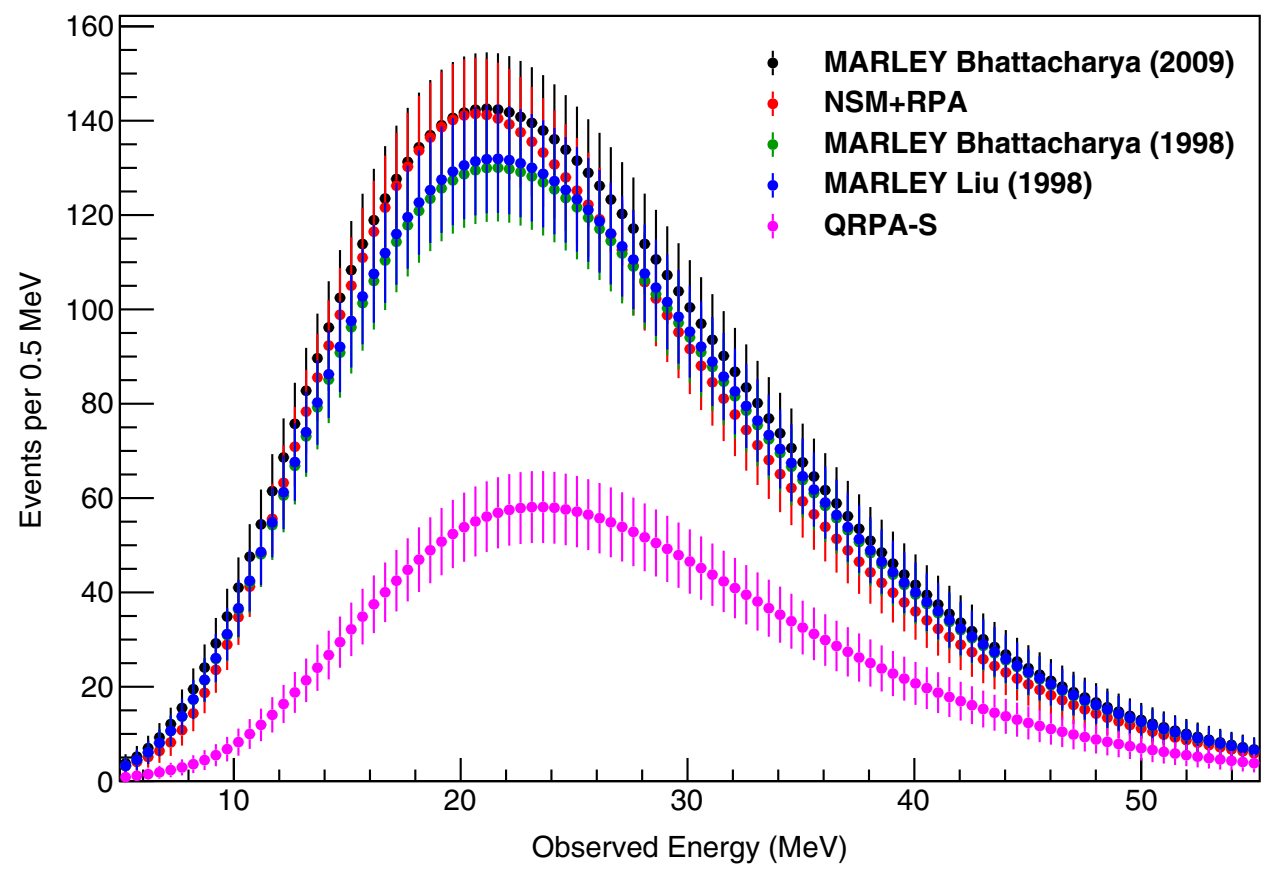

FIG. 16. SNOWGLOBES event rates for the selected cross-section calculations discussed in the text. The error bars are statistical.

With a restricted range of cross-section variations defined in this way, we repeated our fake data studies using a new family of toy cross section models. The lower (Min) and upper (Max) bounds of the uncertainty envelope were treated as two of the new models, and the MARLEY B 2009 cross section [32] was treated as a midpoint. We further define four additional toy models in which three of the models attempt to cover the lower half of the envelope. The first toy model ("Lower bound toy model 1") is an average between the MARLEY B 2009 cross section and the lower (Min) bound. The second toy model ("Lower bound toy model 2") is defined as the average between the first toy model and the MARLEY B 2009 cross section. Finally, the third toy model ("Lower bound toy model 3") is defined as the average between the first toy model and the lower (Min) bound. The complete set of toy cross-section models is shown in Fig. 17. Note that the two "kinks" in the Min model are artifacts from linear interpolations of the NSM + RPA [41] and QRPA-S [44] models, respectively.

Figure 18 shows the 2D fractional difference plots for the toy cross-section models within the uncertainty envelope. When compared to Fig. 13, the biases are less extreme for all three parameters. Similar to the previous fake data studies, extraction of best-fit values for $\alpha$ and $\langle E_\nu \rangle$ is less affected by cross-section mismodeling while estimation of $\varepsilon$ is impacted the most. Also similar to the previous studies, assuming a cross-section higher than the true one leads to an underestimation of $\varepsilon$. Example sensitivity regions are shown in Fig. 19 using several assumed cross sections for fake data generated using the MARLEY B 2009 model. In this case, the black star represents the true parameter values.

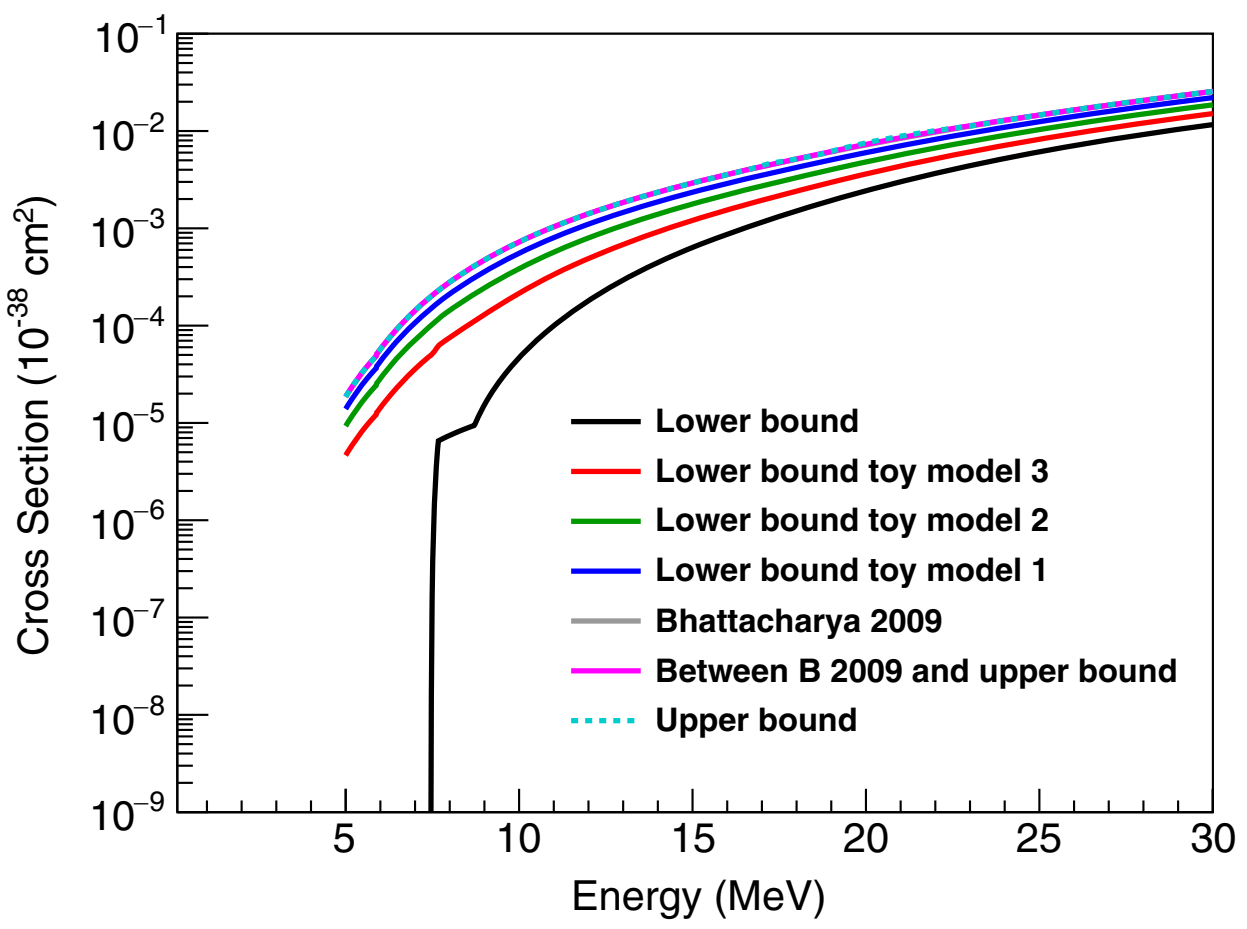

FIG. 17. Toy total cross-section models for the $\nu_e - {}^{40}$Ar interaction covering portions of the uncertainty envelope shown in Fig. 15.

The observed biases are still significant for $\varepsilon$ but relatively modest for the other supernova flux parameters.

## IV. DISCUSSION

A proper interpretation of a DUNE supernova neutrino data set will require a good understanding of neutrino-argon scattering cross sections in the tens of MeV regime. Since direct measurements of the dominant charged-current $\nu_e$ absorption process on argon are currently unavailable, our present consideration of cross-section uncertainties necessarily relies on calculations available in the theoretical literature. Furthermore, because few published calculations of observables beyond energy-dependent total cross sections $\sigma(E_\nu)$ are available for CC $\nu_e - {}^{40}$Ar scattering, we focus entirely upon variations to the total cross section. For the studies reported here, the remaining aspects of the interaction modeling needed to connect the true neutrino energy to the observed energy distribution in DUNE are provided by the MARLEY event generator, which currently implements the only realistic predictions of complete final states for low-energy CC neutrino-argon scattering. We expect the theoretical uncertainties on these additional modeling details to be significant, and future work will be needed to reliably quantify them.

To examine the impact of total cross-section mismodeling on the interpretation of DUNE supernova neutrino data, we employed three strategies for model variations; applying a constant scaling factor to the MARLEY B 2009 model (Sec. III B), considering the full range of a variety of cross-section predictions (Sec. III C), and defining an uncertainty envelope based on the spread of a subset of selected predictions (Sec. III D). Beyond the phenomenological models available in MARLEY, the theoretical calculations that we reviewed and employed for the latter two strategies included the global GTBD treatment and microscopic evaluations such as the QRPA, PQRPA, NSM, and hybrid approaches. All of these models have significant differences coming from the description of nuclear correlations, the residual interaction, and the value of the nucleon axial-vector coupling. Nevertheless, these models reasonably describe the main features of measured weak interaction observables such as $\beta$-decay strengths and inclusive muon capture rates.

For all three strategies, the cross-section model variations were applied to toy measurements of supernova neutrino flux parameters performed using fake data sets produced using the SNOwGLoBES framework. Different combinations of true and assumed cross-section models (used to create the fake data and interpret the toy measurement results, respectively) were employed, and the impact on the extracted values of the flux parameters was assessed.

Table II provides a high-level summary of the conclusions from our fake data studies. For each of the three supernova neutrino flux parameters that we considered, an uncertainty on the total CC neutrino-argon cross-section of $-50/+100\%$ and $\pm 20\%$ is translated into a corresponding range of observed biases on the best-fit parameter value extracted from the toy measurements. The values of the bias were read directly off the 2D fractional difference plots. For the $-50/+100\%$ combination, the forward-fitting algorithm reached the most extreme allowed values of $\varepsilon$, causing the biases in $\alpha$ and $\langle E_\nu \rangle$ to increase in an attempt to compensate for the spectral shape differences between the true spectrum and grid elements.

For total cross section known at about the 20% level, bias on best-fit $\alpha$ and $\langle E_\nu \rangle$ is in the 3–8% range. Achieving less than 10% bias on the best-fit value of $\varepsilon$ requires the cross section to be known to about 5%. These requirements may be somewhat relaxed in light of possible constraints from simultaneous observations of the supernova by other detectors, which we do not consider here. On the other hand, more stringent requirements may ultimately be needed when additional interaction modeling uncertainties

TABLE II. Parameter biases caused by normalization uncertainties on the total cross section.

| $\sigma(E_\nu)$ uncertainty | Parameter | Measurement bias |
|---|---|---|
| $-50/+100\%$ | $\alpha$ | $-80\%$ to $+176\%$ |
| | $\langle E_\nu \rangle$ | $-41.1\%$ to $+47.4\%$ |
| | $\varepsilon$ | $-60\%$ to $+100\%$ |
| $\pm 20\%$ | $\alpha$ | $0\%$ to $+8\%$ |
| | $\langle E_\nu \rangle$ | $-3\%$ to $0\%$ |
| | $\varepsilon$ | $-45\%$ to $+50\%$ |

(beyond those on the total cross section) are fully taken into account.

While we are optimistic that the theoretical understanding of low-energy neutrino-argon cross sections will continue to improve, there is no substitute for actually measuring the cross sections with a well-characterized neutrino flux. Pions decaying at rest represent a near-ideal source of neutrinos for such measurements. Decays of $\pi^+$ produce monochromatic $\nu_\mu$ on a short timescale, plus $\bar{\nu}_\mu$ and $\nu_e$ from delayed decay of the stopped daughter muon on a 2.2 μs timescale. The spectrum and timing are very well understood. The neutrino energies extend to 52 MeV, overlapping nicely with the supernova spectrum. It is also possible to study neutral-current argon inelastic events given the time structure of the beam. Spallation-based neutron beams such as the Spallation Neutron Source at Oak Ridge National Laboratory [53], the Lujan Neutron Science Center at Los Alamos National Laboratory [54], the J-PARC Spallation Neutron Source [55], and the future European Spallation Source [56] (currently under construction) are intense sources of pion decay-at-rest neutrinos. Measurements of these neutrinos may also be possible at high-energy physics facilities including the Large Hadron Collider beam dump [57] and the meson decay-in-flight neutrino beams at Fermilab [58].

Future direct measurements of CC $\nu_e$-argon cross sections using a pion decay-at-rest source could pursue several distinct observables to better constrain interaction modeling uncertainties for the DUNE supernova neutrino program. The most straightforward of these (and most directly relevant to the specific uncertainties considered in this paper) would be an inclusive total cross section $\langle \sigma \rangle$ averaged over the $\nu_e$ flux $\phi(E_\nu)$ from $\pi^+$ decays at rest,

$$\langle \sigma \rangle \equiv \frac{\int_0^{m_\mu/2} \sigma(E_\nu)\phi(E_\nu)dE_\nu}{\int_0^{m_\mu/2} \phi(E_\nu)dE_\nu}, \tag{9}$$

where $m_\mu$ is the muon mass and

$$\phi(E_\nu) \propto E_\nu^2 m_\mu^{-4}(m_\mu - 2E_\nu). \tag{10}$$

Measurements of both $\langle \sigma \rangle$ and a differential cross section as a function of the total visible energy would likely be obtainable with a suitably large (several-ton-scale) argon detector. As an example, 5–10% statistical uncertainty on the total cross section could be obtained in a few years with a ton-scale detector a few tens of meters from the Spallation Neutron Source.

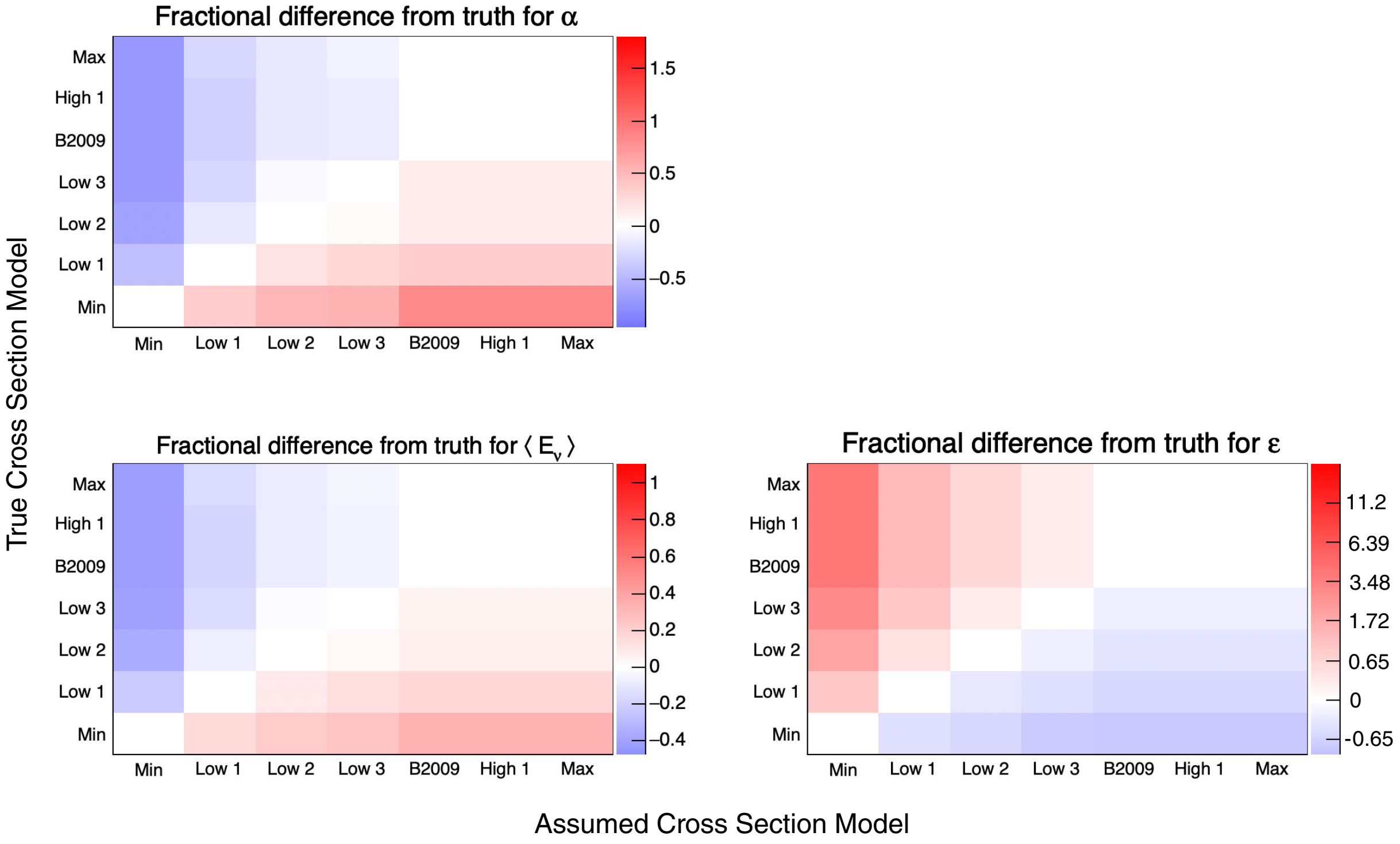

FIG. 18. 2D fractional difference plots to study effects produced by toy models within the cross-section uncertainty envelope discussed in Sec. III D.

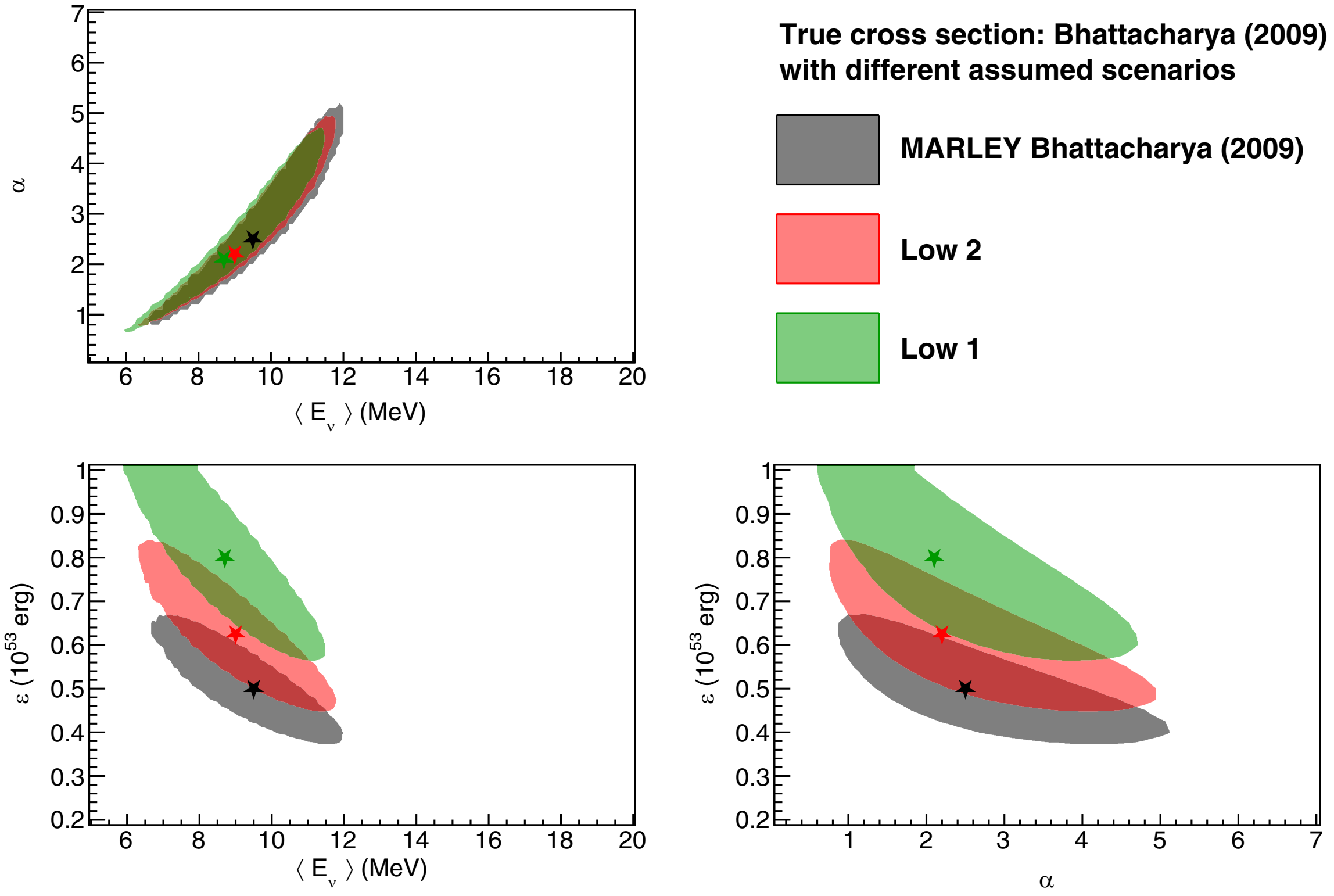

FIG. 19. Sensitivity regions (90% C.L.) with different combinations of assumed and true cross-section models. Two of the models are toy models generated from the midpoint (B 2009) and minimum cross-section values from the set of selected models. The stars mark the best-fit values from the fitting algorithm. The black stars also represent the true parameter values.

The fine spatial resolution of a LArTPC detector would potentially allow for more detailed measurements. In particular, topological separation between the outgoing electron and $\gamma$-rays emitted due to neutrino-induced nuclear deexcitations could allow separate measurements of differential distributions for both particle species. Recent studies (e.g., Ref. [59]) suggest that such a separation would be feasible, and a successful implementation would yield a rich data set; the inclusive electron energy and angular distributions are known to be sensitive to the modeling of forbidden contributions to the cross section [60], while the $\gamma$-rays would provide a helpful constraint on deexcitation modeling and, in principle, the opportunity to measure partial cross sections for specific nuclear transitions. Measuring the neutrino angular distribution is particularly important for supernova pointing measurements relevant for prompt multimessenger astrophysics [12,61].

An especially impactful but highly challenging measurement would involve the detection of final-state neutrons produced by CC $\nu_e$-argon interactions. Missing energy attributable to these neutrons is expected to have a significant impact on neutrino energy reconstruction at supernova energies [31], and the modeling needed to account for it is complicated and poorly constrained by experimental data. In the absence of any new experimental techniques to increase the sensitivity of argon-based detectors to neutrons at and below MeV energies, external instrumentation designed to capture and detect escaping neutrons would likely be the only means of attempting such a measurement.

## V. CONCLUSION

A possible future observation by DUNE of neutrinos from a core-collapse supernova would represent a rare and valuable scientific opportunity. In particular, the unique sensitivity of DUNE's LArTPC detectors to the $\nu_e$ component of the supernova neutrino flux would be highly complementary to other current and anticipated large neutrino experiments. In the studies reported in this paper, we have examined the effects of cross-section modeling uncertainties on a simulated analysis of supernova neutrinos in DUNE.

Significant experimental and theoretical challenges remain before a precise understanding of tens of MeV neutrino-argon scattering can be achieved. Nevertheless, pursuing this understanding will be essential to maximize the discovery potential from a core-collapse supernova observation (and a potentially broader program of low-energy physics) in DUNE. We hope that the initial studies of neutrino-argon interaction modeling uncertainties reported here may serve as a useful foundation for the more comprehensive investigations that will be required in the future.

## ACKNOWLEDGMENTS

This document was prepared by the DUNE Collaboration using the resources of the Fermi National Accelerator Laboratory (Fermilab), a U.S. Department of Energy, Office of Science, HEP User Facility. Fermilab is managed by Fermi Research Alliance, LLC (FRA), acting under Contract No. DE-AC02-07CH11359. This work was supported by CNPq, FAPERJ, FAPEG and FAPESP, Brazil; CFI, IPP and NSERC, Canada; CERN; MŠMT, Czech Republic; ERDF, H2020-EU and MSCA, European Union; CNRS/IN2P3 and CEA, France; INFN, Italy; FCT, Portugal; NRF, South Korea; CAM, Fundación "La Caixa," Junta de Andalucía-FEDER, MICINN, and Xunta de Galicia, Spain; SERI and SNSF, Switzerland; TÜBİTAK, Turkey; The Royal Society and UKRI/STFC, United Kingdom; DOE and NSF, United States of America. This work was also supported by FAPESB T. O. PIE 0013/2016 and UESC/PROPP 0010299-61.

## APPENDIX A: INTERPOLATION/ EXTRAPOLATION METHODS USED ON CROSS SECTION MODELS

In order to study the measurement biases introduced by the cross-section modeling, we obtained numerical tables of model predictions for the total charged-current $\nu_e - {}^{40}\mathrm{Ar}$ cross section (see Table I). SNOwGLoBES requires 1001 data points in a cross-section file for neutrino energies between 5–100 MeV. While some of the models of interest are already available within SNOwGLoBES (including its default cross-section model, along with some MARLEY cross-section models from Ref. [31]), input files for the other models required extra preparation to conform to the requirements of the SNOwGLoBES format.

Table III summarizes the interpolation and extrapolation methods used for the various models. Excluding the cross-section models already available within SNOwGLoBES, all models required interpolation between their tabulated data points to obtain cross-section values at intermediate neutrino energies. For models which were tabulated over the entire energy range of interest, either a cubic spline or a linear spline was used to interpolate between the given data points. A cubic spline was generally preferred, but the linear spline was used in cases where the cubic spline caused unphysical fluctuations in the interpolated total cross section.

The available cross-section tables for some models did not cover the entire 5–100 MeV energy range required by SNOwGLoBES. In such cases, extrapolation techniques were used to extend the existing predictions. The models from Refs. [38,39,41,45,46] required extrapolation down to 5 MeV, while the model from Ref. [43] required extrapolation down to 5 MeV and up to 100 MeV. All of the extrapolations used to prepare the SNOwGLoBES input files employed a quadratic fit of the form

$$\sigma(E_\nu) = p_0(E_\nu - p_1)^2, \tag{A1}$$

where $p_0$ and $p_1$ are the free parameters used for fitting. All extrapolation fits used five data points.

TABLE III. Table summarizing the interpolation and extrapolation methods performed on the various cross-section models to format them for usage in SNOWGLOBES [29]. Parameters from the quadratic fits described in the text are also given when extrapolation was used.

| Cross-section model | Interpolation method used | Extrapolation method used |
|---|---|---|
| SNOWGLOBES [29] | Not applicable | Not applicable |
| RPA [38,39] | Linear spline | Low-energy quadratic fit: $\sigma = 1.35027 \times 10^{-5}\,(E - 0.567063)^2$ |
| QRPA-C [43] | Linear spline | Low-energy quadratic fit: $\sigma = 7.29830 \times 10^{-6}\,(E - 6.67699)^2$; for all energy values below $p_1 = 6.68$ MeV, the cross section was set to zero. High-energy quadratic fit: $\sigma = 1.83273 \times 10^{-5}\,(E - 12.3510)^2$ |
| GTBD [45,46] | Linear spline | Low-energy quadratic fit: $\sigma = 2.26358 \times 10^{-5}\,(E + 0.761242)^2$ |
| NSM + RPA [41] | Linear spline | Low-energy quadratic fit: $\sigma = 1.49812 \times 10^{-4}\,(E - 7.45969)^2$; for all energy values below $p_1 = 7.46$ MeV, the cross section was set to zero. |
| QRPA-S [44] | Linear spline | Not applicable |
| RQRPA [40] | Cubic spline | Not applicable |
| PQRPA [42] | Cubic spline | Not applicable |
| B 1998 [32] | Cubic spline | Not applicable |
| B 2009 [32] | Cubic spline | Not applicable |
| L 2009 [32] | Cubic spline | Not applicable |

In the fits for low energies, $p_1$ (which has units of MeV) holds special significance as the "endpoint" of the cross-section model because it is the minimum of the quadratic function. For $p_1 > 5$ MeV, the fit would introduce unphysical behavior into the model in the form of an increasing cross section as the neutrino energy $E_\nu$ approaches 5 MeV from above. To prevent this behavior, the total cross section $\sigma(E_\nu)$ was zeroed out for all energies $E_\nu < p_1$ whenever $p_1 > 5$ MeV. The same quadratic functional form was also fit to the last five data points of the model from Ref. [43] to extrapolate up to 100 MeV. In this case, the low- and high-energy fits were handled independently. In order to avoid discontinuities between the interpolation and extrapolation methods, the fits performed at low (high) neutrino energy were required to pass through the first (last) tabulated data point for the cross-section model of interest. Figure 20 shows the cross section model from Refs. [45,46] as an example of the interpolation between points (in this case, with a linear spline) as well as an extrapolation to low energies.

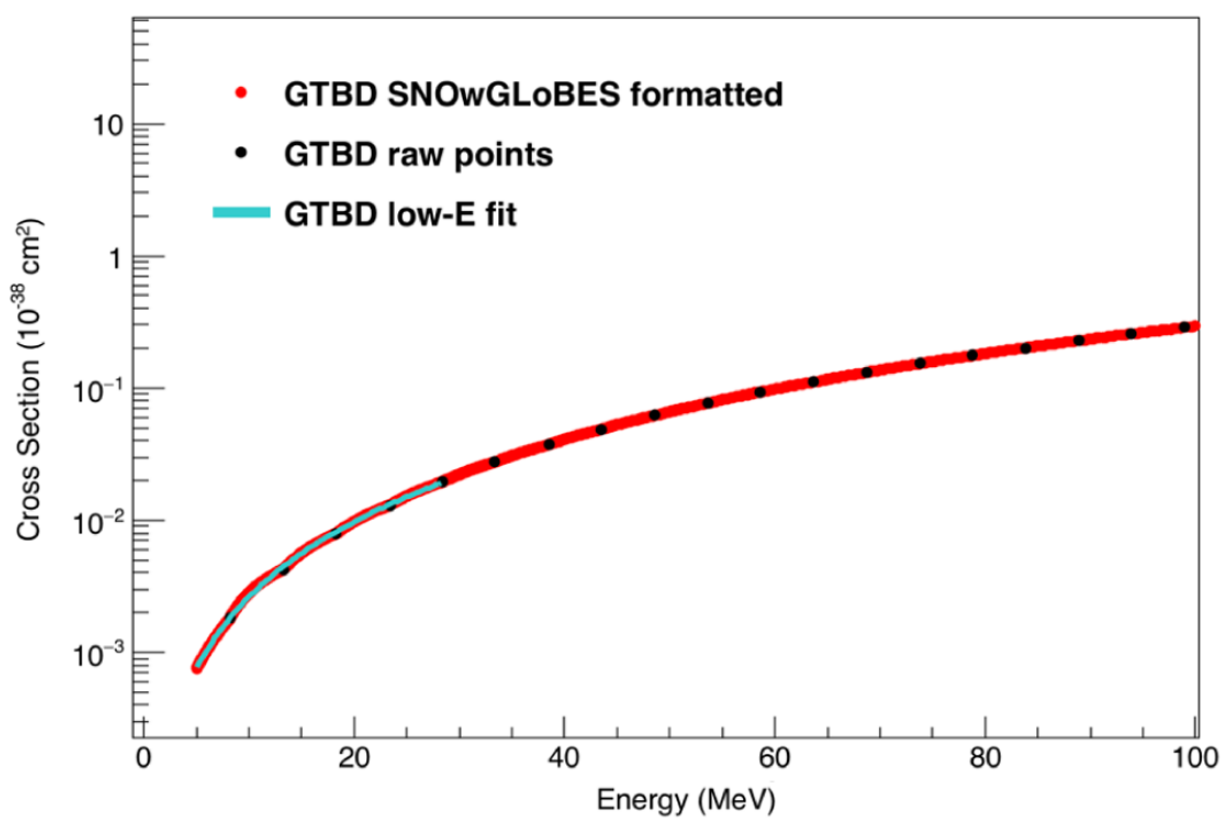

FIG. 20. Cross-section model from Refs. [45,46] with the interpolation (with a linear spline) and extrapolation (using a quadratic fit) shown. See Table III for the quadratic fit parameters for the low-energy fit.

## APPENDIX B: SNOwGLoBES EVENT RATES FOR DIFFERENT CROSS-SECTION MODELS

TABLE IV. SNOWGLOBES estimated number of $\nu_e$CC events in the DUNE far detectors for pinched-thermal flux parameters $(\alpha, \langle E_\nu \rangle, \varepsilon) = (2.5, 9.5, 5 \times 10^{52})$ for the $\nu_e$ flavor, a 10 kpc supernova, and assuming NMO and MSW oscillations via Eq. (5).

| Cross-section model | Number of $\nu_e$CC events | Number of $\nu_e$CC events between [5, 15] MeV |
|---|---|---|
| QRPA-C [43] | 1383 | 134 |
| RQRPA [40] | 2243 | 220 |
| QRPA-S [44] | 2791 | 243 |
| SNOWGLOBES [29] | 4486 | 624 |
| B 1998 [32] | 6307 | 874 |
| L 1998 [32] | 6390 | 883 |
| NSM + RPA [41] | 6391 | 897 |
| B 2009 [32] | 6852 | 988 |
| PQRPA [42] | 4562 | 909 |
| RPA [38,39] | 5064 | 998 |
| GTBD [45,46] | 7770 | 2070 |

## APPENDIX C: INTERPOLATING SENSITIVITY REGIONS

To keep computation time reasonable, the algorithm used to compute flux parameter sensitivity regions (see Sec. II D) uses a limited number of elements in the grid of reference $(\alpha, \langle E_\nu \rangle, \varepsilon)$ values. The limited number of grid elements leads to unphysical jagged edges in plots of the 90% confidence contours used in this paper to estimate DUNE sensitivity regions for the supernova spectral parameters. To remove these artifacts from the sensitivity region plots, we developed an interpolation technique to smooth the contour edges. Each contour was stored as a two-dimensional histogram, where the weight in each bin was calculated as the minimum $\chi^2$ value obtained in that region of 2D flux parameter space. Bilinear interpolation [62] between histogram bins was then used to increase the number of bins along each axis to 1000. Example sensitivity regions for the MARLEY B 2009 model are shown in Fig. 21 before (black) and after (blue) applying the smoothing procedure. The impact of the smoothing is most noticeable in the plots involving $\varepsilon$ since the reference grid is coarsest for that parameter. Specifically, the interpolated contours are slightly smaller than the original contours.

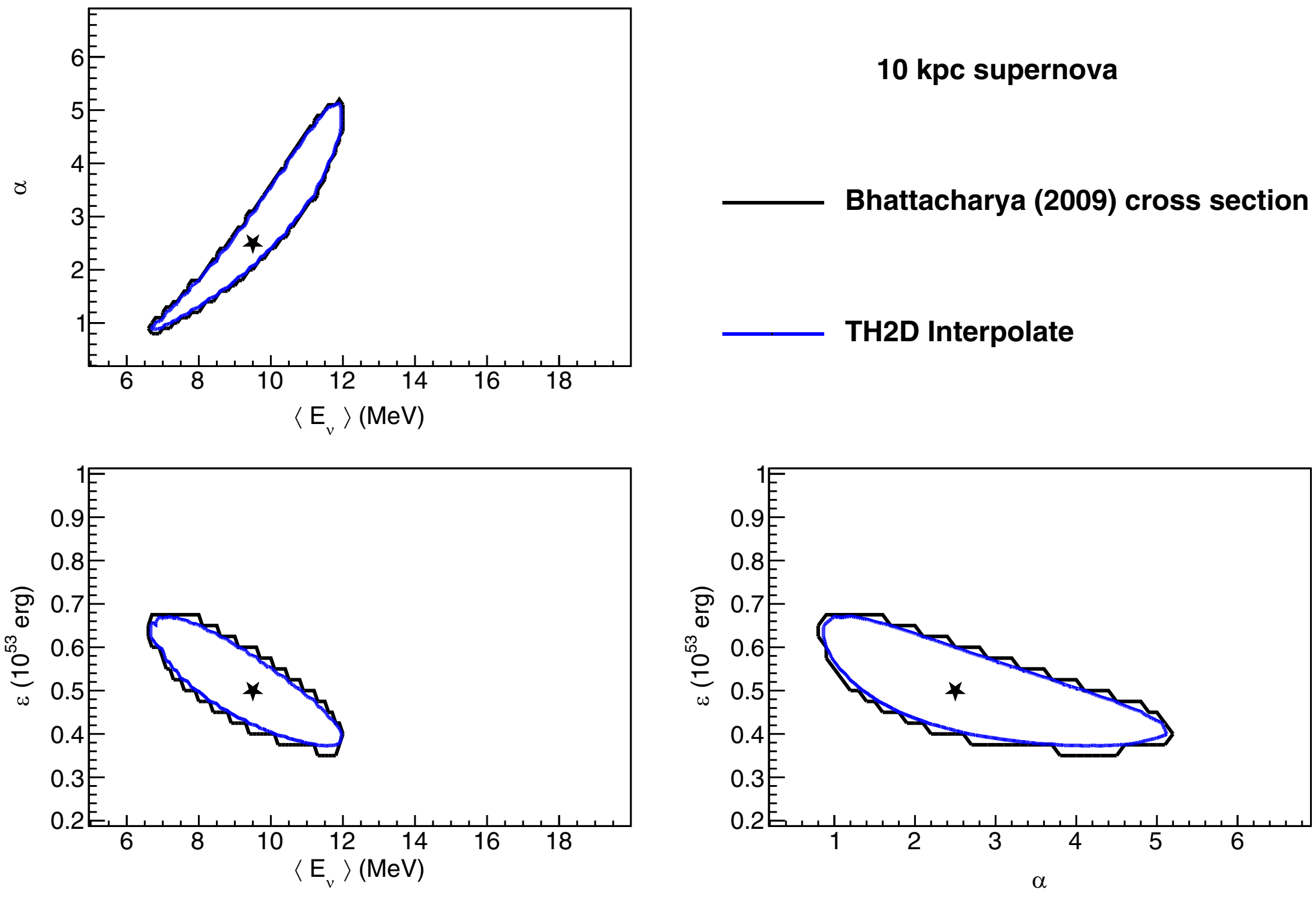

FIG. 21. 90% C.L. contours for the three parameter spaces with NMO assumptions and the MARLEY B 2009 cross-section model [32]. The contours before interpolation have prominent jagged edges due to a limited number of reference grid points. The edges are most noticeable for the $\varepsilon$ parameter.